\documentclass[journal]{IEEEtran}
\usepackage{ifpdf}
\usepackage[pdftex]{graphicx}
\usepackage{subfigure}
\usepackage{amsmath}
\usepackage{algorithmic}
\usepackage{array}
\usepackage{stfloats}
\usepackage{color}
\usepackage{soul,color}
\usepackage{cite}
\usepackage{booktabs}
\usepackage{float}
\usepackage{makecell}
\usepackage{diagbox}
\usepackage{multirow}
\usepackage{inputenc}   
\usepackage{upgreek}
\usepackage{url}
\usepackage{comment}
\usepackage{makecell}
\usepackage{textcomp}

\begin{document}

\title{A Highly-Compact Direct-Injection Universal Power Flow and Quality Control Circuit}

\author{

        Mowei Lu,~\IEEEmembership{Student Member,~IEEE,}
        Mengjie Qin,~\IEEEmembership{Student Member,~IEEE,}
        Jan Kacetl,~\IEEEmembership{Student Member,~IEEE,}
        Eeshta Suresh,~\IEEEmembership{Student Member,~IEEE,}
        Teng Long,~\IEEEmembership{Member,~IEEE,}
        Stefan M. Goetz,~\IEEEmembership{Member,~IEEE,}

        }

\markboth{IEEE~JOURNAL~OF~EMERGING~AND~SELECTED~TOPICS~IN~POWER~ELECTRONICS,~Vol.~XX,~No.~XX,XXXX~20XX}{Xuejiao Pan \MakeLowercase{\textit{et al.}}: Harmonic cancellation by Adaptive notch filter based on Wavelet Package Transform for an MMCC-STATCOM}

\maketitle

\begin{abstract}

This paper presents a novel direct-injection modular
universal power flow and quality control topology exclusively
using lower power components. In addition to conventional
high-voltage applications, it is particularly attractive for the
distribution and secondary grids, e.g., in soft open points, down
to low voltage as it can exploit the latest developments in
low-voltage high-current semiconductors. In contrast to other
concepts that do not interface the grid through transformers,
it does not need to convert the entire line power but only the
injected or extracted power difference. The proposed power flow
and quality (f/q) controller comprises a shunt active front end,
together with high-frequency links serving as a power supply
for a series floating module per phase. Each of the floating
modules is in series with one phase of the line, floating with
the electric potential of that particular phase, avoiding any
ground connection. Omitting bulky and dynamically limited line
transformers of conventional universal power flow controllers,
the presented direct-injection f/q controller enables exceptionally
small size and volume, high power density, high frequency
content, and fast response. In contrast to direct-injection concepts
with full back-to-back converters, it only needs to handle a
fraction of the power. The circuit combines grid-voltage lowcurrent
electronics in the shunt unit and low-voltage high-current
modules in the floating series injection units. Simulations and
experiments demonstrate and validate the concept.
\end{abstract}


\section{Introduction}

\IEEEPARstart{T}{he}  development of dedicated power electronics technology has increased the flexibility and capacity of the higher-voltage grid levels \cite{Understanding_FACTS,High_Wire_Act,Global_Energy_Scenario}. Compensation and power-flow control units can absorb reactive power and distortion, increase the active capacitance, control currents, and reduce voltage range violations \cite{New_approach}. The low-voltage grid, however, mostly consists of conventional mechanical or passive components.
{At best, the low-voltage grid has slow on-load tap changers (OLTC) and switched shunt capacitor banks (CB). The overall grid is facing two major developments, which culminate in the low voltage: (1) a large level of widely uncontrolled generation in residential buildings and (2) the electronification of previously rather slow and forgiving devices which now feature active feedback control and switched-mode operation.} 

{In conventional grids, the power flow was primarily unidirectional and well-defined from large power plants feeding the higher voltage levels to the households and most businesses in the low-voltage grid. The low voltage grid was dominated by relatively small power units; drives, heating, and similarly tolerant loads were usually the only larger power consumers. In the meantime, increasing distributed renewable power generation in residential buildings feeds power into the low-voltage grid without any central control \cite{Electric_Vehicle,PV_JESTPE}. In addition, more and higher-power electronic sources and loads in switched mode with rapid control dynamics, such as vehicle chargers, can introduce distortion and rapid load changes \cite{On_the_Stability,9940564,An_Estimation_Based}.
}

As a consequence of these developments, undesired power flows, which can be nonobvious in meshed grids, may overload lines and let the voltage locally leave the typical ±5\% or even ±10\% tolerance band, sometimes in both directions in the same grid segment \cite{A_novel_impact_assessment}. { In the presence of distributed power generation and loads, adjustable transformers---neither with on-load tap changers nor electronic---can typically not solve this situation. They can only simultaneously increase or decrease the voltage of the entire feeder; however, the voltage differences in loops, which drive the unwanted power flows, do not change \cite{LV_Grid_Tap_Changing,Parallel_Operation_Transformers,Biela}.
Distortion and instabilities of power electronics on the low-voltage level can destabilize and congest the grid not only locally but spread to the medium-voltage grid.}
The flexible ac transmission system (FACTS) catalogue includes unified power flow controllers (UPFC) and static compensators (STATCOM) to solve related problems on the high-voltage level \cite{9146142,Biela,8960648}.

In the low-voltage grid, unified power quality conditioners (UPQC) and the introduction of controllable interconnections, so-called soft open points (SOP), are intensively studied as solutions \cite{7132779,Operating_principle_of_Soft_Open_Points,Optimal_operation_of_soft_open_points,Soft_Open_Point_in_Distribution_Networks}. SOPs can establish a connection between feeders in the distribution grid and actively control the power flow and/or the voltage difference between both. { SOP circuits can be implemented in various ways. Most obviously and widely promoted, SOP circuits use full back-to-back converters, which convert the entire power flowing across from the ac of one feeder to a dc link and back to ac of the other one. Alternatively, they employ low-voltage UPFC  as well as related UPQC variants that go beyond filtering \cite{An_Overview_of_Soft_Open_Points,Increasing_dis_generation_pene,7132779}.} 

UPFC and UPQC circuits share similar topologies. Many UPFC circuits can perform UPQC functionality and vice versa although some topologies may be limited, e.g., in power injection or compensation of higher-frequency distortion. { The traditional configuration consists of a common dc link, a shunt voltage source converter, and a series voltage source converter, coupling to the transmission line through injection transformers with a high-current secondary winding \cite{6226856,LVDN}}.
The converters can, for instance, be based on modular multilevel converters and similar cascaded structures for modularity and reliability \cite{8003361,9216052}. { Alternatively, matrix converters or direct power flow controllers can replace the dc link with a direct ac–ac connection \cite{Matrix,M3C_Based,DirectPFC}}. {However, the bulky series-injection transformer is an essential element of most known solutions and often dominates units with respect to size. Thus, the transformer typically causes large installations for sufficient thermal and magnetic rating for typical line currents, additional losses, as well as limited dynamical response \cite{UPFC_transformer, Lu_Transformerless, Solid_transformer}}.
 Due to magnetic loss and parasitic { resonance}, transformers limit the possible injection to only a narrow frequency range \cite{7448961}. Recent research aims at eliminating transformers through full back-to-back converters and/or high-frequency links \cite{TransformerlessUPFC,Transformer_Less_UPFC_two,UPFC_transformer,Hybrid_UPQC}. However, most solutions suffer from the requirement of handling both the full line voltage and current. Various cunning
concepts allow a reduction of the number of semiconductors,
but each semiconductor still has to be a costly high-power
device \cite{Transformer_Less_LED,UPQC_Remote_Sytem}.

{ This paper proposes a novel direct-injection circuit for UPFC, UPQC, and SOP functions, and provides more detailed operating principles and validation based on our previous concept \cite{IECON}}. { In contrast to other direct-injection designs that lack the ability to exchange active power with the grid, this circuit is capable of compensating active/reactive power, controlling power flow amplitude/direction, and filtering harmonic content \cite{TransformerlessUPFC,Transformer_Less_UPFC_two,DVC}}. Although it can also be applied to high-voltage grids, it presents unique advantages for meshed low-voltage grids when implemented with the latest low-voltage high-current semiconductors. By eliminating low-frequency transformers connected in series or parallel with the grid, the circuit is able to achieve high power density and superior dynamics.

\begin{figure*}[t]
    \centering
    \includegraphics[width=0.8\linewidth]{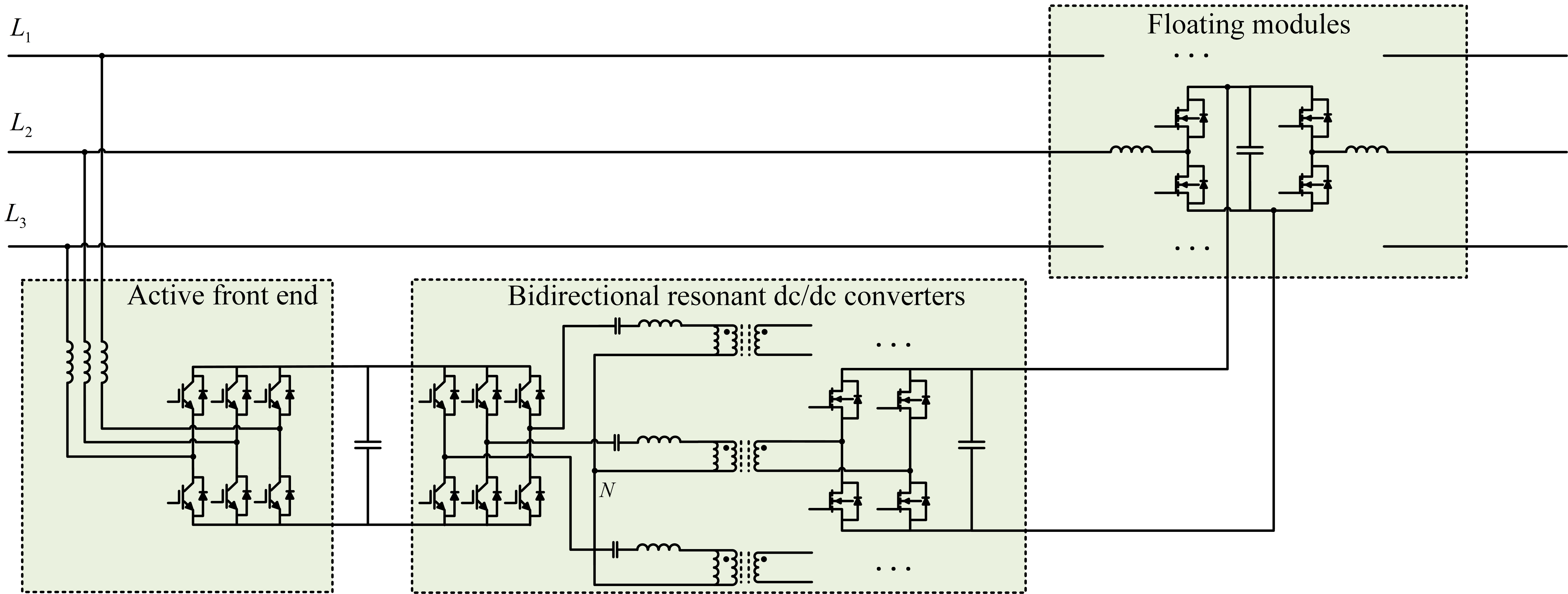}
    \caption{Circuit configuration of the DI-UPF/QC.}
    \label{fig:DWPT}
\end{figure*}

The system configuration of Fig.\ \ref{fig:DWPT} uses a full-grid-voltage low-current power supply containing a { shunt active front end (AFE)}, a bidirectional full-bridge resonant dc/dc converter per phase shifting the supply power to higher frequencies for compactness and less magnetic material in the isolation stage, and a floating high-current low-voltage series module. The design leverages on our previous work on modular floating, high-voltage low-current, and low-voltage high-current circuits, thus exploiting the development of power semiconductors of recent years driven by consumer electronics, appliances, and the automotive industry to achieve enormous power density and dynamics \cite{Review_MMC,GOETZ2019667,GoetzMMC,LiTMS,Goetz2012}. { Since the circuit can use existing higher-level control techniques and functions from SOPs and UPQCs, the original contributions of this design focus on the topology with the following target:
}

(1) {Compared to transformer-based systems, the direct-injection f/q controller has a fast response as well as wide bandwidth and is small in size.}

(2) { Compared to full back-back converters, the circuit only needs to convert a fraction of the line power. With a small voltage injection, it can push tremendous currents, even short-circuit currents in the \textgreater\,{}1,000~A range. The comparably low line impedance in the grid generates substantial leverage and requires only low voltages for controlling the flow, as low as a few volts for the low-voltage grid. As the series-injection modules ride on the phase voltage, they can exclusively use low-voltage components.}

(3)	{ Compared to conventional UPFCs, which typically aim at point-to-point transmission connections, the presented circuit is a new concept particularly advantageous to control power flow in lower-voltage grid levels. These lower-voltage grid levels are experiencing increasing difficulties with dynamic high-power loads, such as vehicle chargers, a high share of (often constant-power-regulated) power electronics, and distributed renewable generation, such as residential solar power.}

(4) The circuit combines high-voltage low-current electronics in the AFE and low-voltage high-current components in the floating modules for a highly dynamic compact direct-injection system. 
In contrast to semiconductor devices with full voltage and current rating as needed for back-to-back converters, these components with only either high current or high voltage are readily available and can switch rather fast.

\section{Operation Principles}

\subsection{System Overview}

Due to the relatively low impedance of grids (even weak grids), already a small voltage difference injected into a grid mesh or two grid segments is sufficient to drive large currents and powers. For low-voltage distribution grids, sufficient driving voltages can be as low as 5 -- 10~V \cite{Transformer_Less_UPFC_two,An_Estimation_Based,Grid_Impedance_Estimation}. The entire grid current has to be lifted or reduced by this voltage similar to the water flow in a canal lock though. These currents can reach hundreds of amperes under normal conditions and kiloamperes during faults for the example of a low-voltage grid. Without any ground connection, the floating modules only need to generate and control the small voltage difference and the current (Fig.\ 1). {The power actively injected or extracted is low to moderate for power-flow control, while power-factor correction and harmonics compensation do practically not involve any active power beyond compensating losses and are provided by the dc link. Thus, the power of the supply and shunt circuit will need to connect to the grid voltage like any simple conventional power supply, but the current is small.}

Consequently and as outlined above, the circuit contains two parts: an isolated power supply with parallel grid interface (shunt part) and a floating series injection module. To enable the modules to control power flow from a higher-voltage to a lower-voltage grid in a four-quadrant operation, a bidirectional power supply is necessary. {The proposed solution is to use a shared active front end (AFE) and a shared dc link, along with a single bidirectional, preferably resonant dc/dc converter per module for isolation and to reduce the voltage of the shared dc link to the module supply voltage level. }
As no tight voltage regulation is needed between the module and the shared dc link on the AFE side, an open-loop-operated { LLC (inductor--inductor--capacitor) resonant converter suffices.} Other isolated dc/dc converters can likewise serve for the power exchange. { The AFE with appropriate grid filter can serve as a highly dynamic shunt converter following known concepts \cite{PWM_regenerative,LLCL,LLCL2,10.1049/pel2.12415}}. Moreover, it absorbs or supplies active and reactive power to the dc link of the floating module. The shared dc link behind the AFE allows different asymmetric load conditions on the AFE and the module side while power can be exchanged between the phases. 

Instead of using low-frequency bulky transformers, the LLCs isolate the electrical dc-link potential from the floating modules and shift the voltage with a fixed voltage ratio from the grid-level higher voltage to the module-level lower voltage. The isolation allows the module to float with the line voltage. While any isolating bidirectional dc/dc topology can be used, an LLC provides high efficiency when in resonance, generates low electromagnetic interference, and maintains a relatively constant voltage ratio without the need for delicate control of the dc-link voltage \cite{BidiLLC}. This makes it possible to use an open-loop control method.

The centrepiece of the system is a floating module connected in series with each grid phase. Each module consists of a dc capacitance, a transistor H bridge, and at least one grid inductor. For symmetric, inductors are used on both sides, as shown in Fig.~\ref{fig:DWPT}. Since the floating modules follow the electrical potential of their assigned phase and have no connection to ground or the other phases, the local voltages are low, e.g., below 48~V for low-voltage grids. As a result, the modules are considered low-voltage units and can employ low-voltage capacitors and transistors. Driven by automotive and consumer electronics, latest field-effect transistors (FET) offer high current capabilities below 100~V \cite{LV_HC,Trench_Power_MOS,Trench_Power_MOS_2}.

\begin{figure}[h]
    \centering
    \includegraphics[width=1\linewidth]{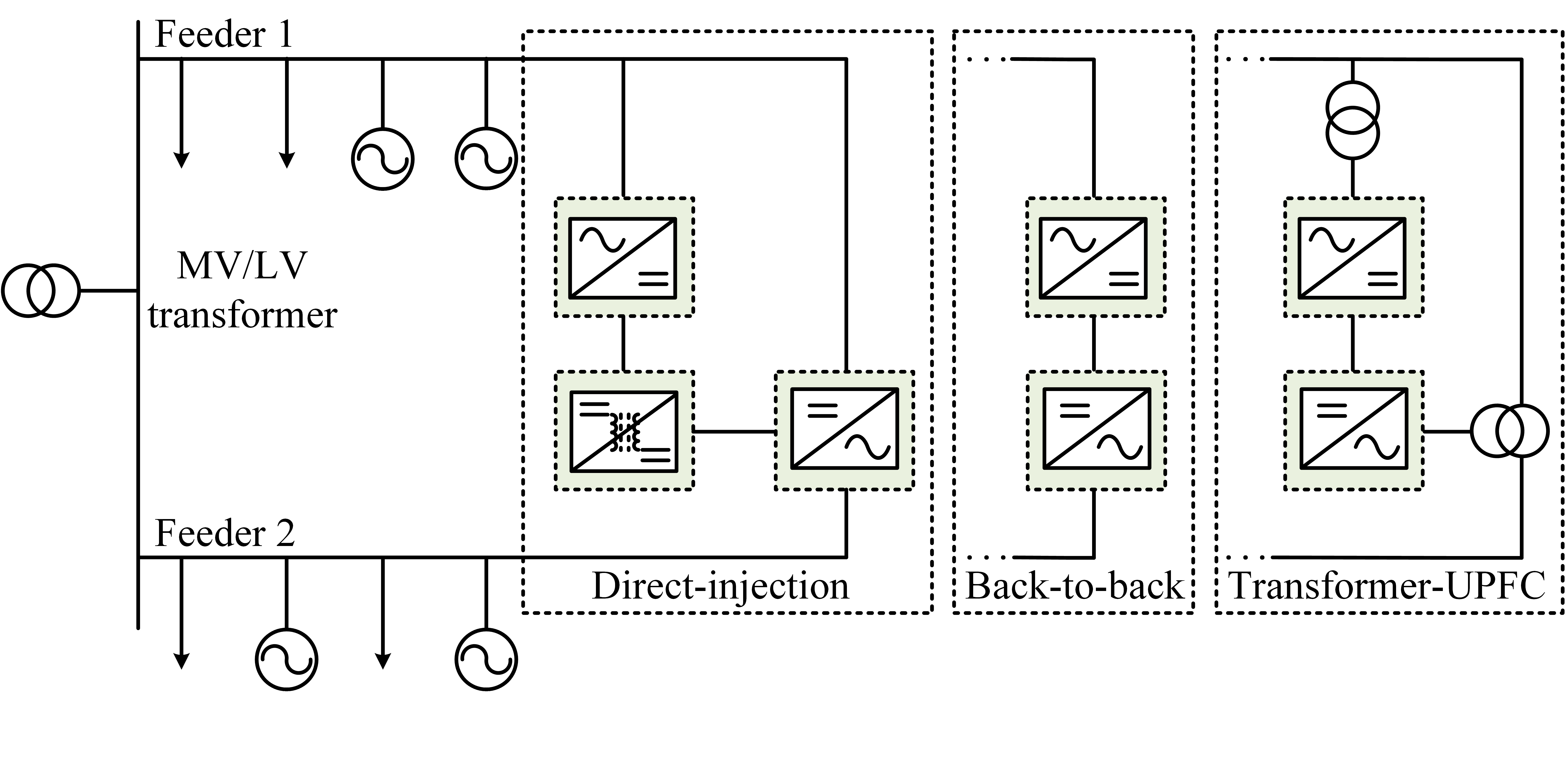}
    \caption{{Different power flow and quality control topologies in low-voltage distribution grids.}}
    \label{fig:sys}
\end{figure}

{  ${L_{\rm{1}}}$, ${L_{\rm{2}}}$, and ${L_{\rm{3}}}$ in Fig.~\ref{fig:DWPT} are cables or crossbars between two ac distribution networks (feeders or buses), in a loop in a grid, or between a grid and a load. Fig.~\ref{fig:sys} illustrates the deployment of the proposed direct-injection circuit between two feeders, alongside a comparison to the configurations involving back-to-back converters and transformer-based UPFC. Each of the previous solutions has specific disadvantages, e.g., full power need for the back-to-back converter or a large injection transformer for the UPFC. The proposed topology avoids the individual problems of each approach but maintains their advantages and combines them. Just as back-to-back converters, the power electronics directly connect to the grid. Similar to conventional UPFCs, the circuit only does not need to convert the entire controlled power.}  

The modules in the proposed design are able to control large power flows using only a fraction of the power required by injecting or extracting the necessary difference voltage and associated power. { Since there are no line-frequency injection transformers involved, the electronic interface allows for current and voltage control with a wide bandwidth into the kilohertz range, similar to low-voltage inverters.} Despite handling the full grid current, the modules require only a fraction of the voltage needed to control power flow, phase shift, or inject/absorb harmonics. { The dc/dc converters just need to provide the power injected into the grid, or remove the power extracted from the grid, at the module dc voltage, such as 48~V.} The AFE stage, which converts the net real power left after the shared dc link capacitor, absorbs reactive power from the three floating modules and active power flows between them. The AFE component is not new and can follow known topologies and control principles from the prior art.

{
\subsection{AFE control}
The overall target for the active front-end (AFE) control is to obtain a steady dc voltage that can provide power to the floating module after it has been stepped down by the subsequent high-frequency link. In the low-voltage grid, phase asymmetries are common since single-phase loads are usually unevenly distributed, and the grid itself normally operates in an unbalanced manner. This can result in ripple and also even-order harmonics at the dc link. To mitigate these issues, the control of the AFE can use well-established dual current control to eliminate the negative sequence \cite{Unbalanced}.
\begin{figure}[h]
    \centering
    \includegraphics[width=0.5\linewidth]{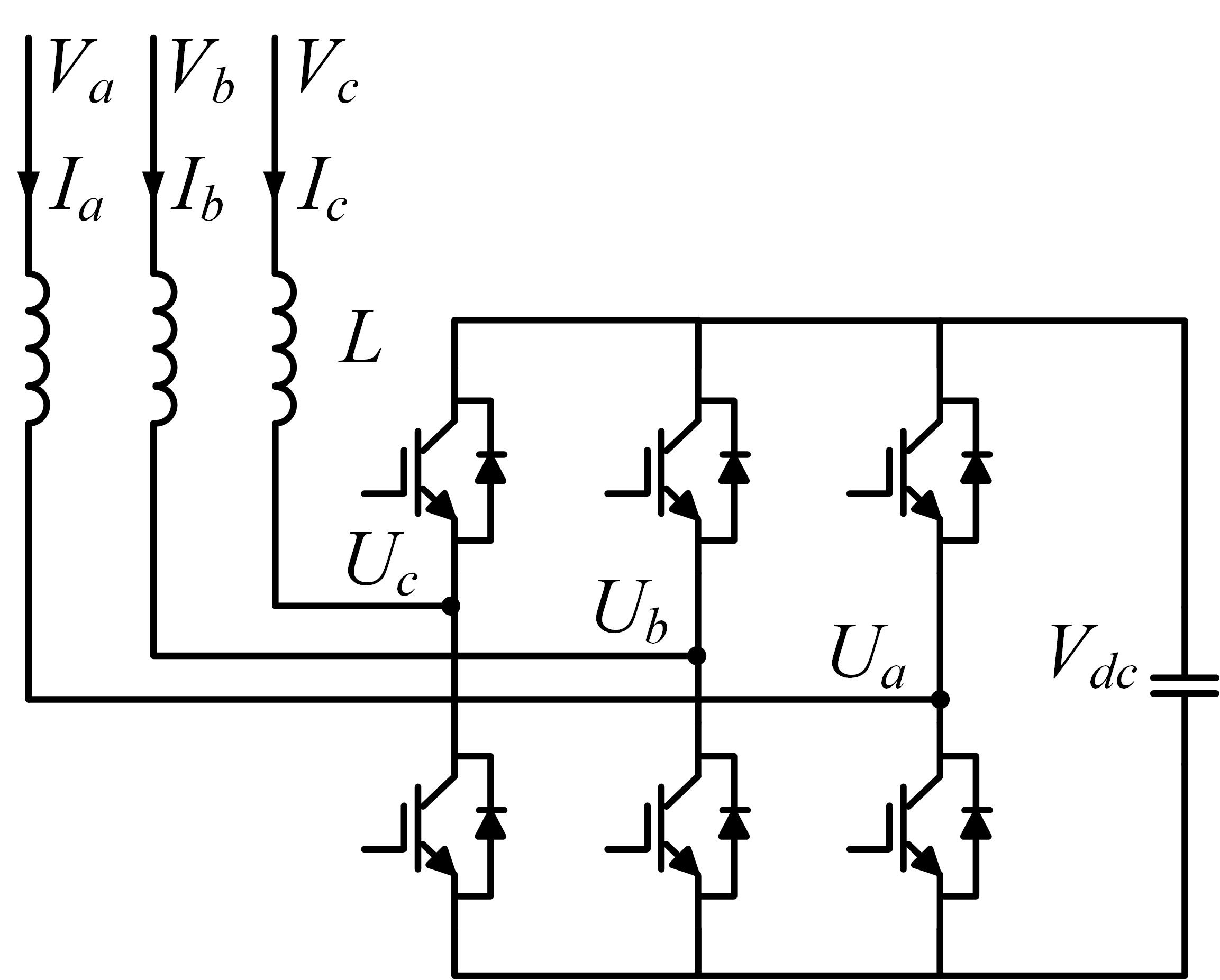}
    \caption{The equivalent circuit of the active shunt front end.}
    \label{fig:AFE}
\end{figure}

Fig.~\ref{fig:AFE} displays the equivalent circuit of the AFE. The positive-sequence node voltage follows
\begin{align}
\left\{ {\begin{array}{*{20}{c}}
{U_{\rm{d}}^{\rm{p}} = V_{\rm{d}}^{\rm{p}} - ({k_{\rm{p}}} + \frac{{{k_{\rm{i}}}}}{s})(I_{\rm{d}}^{{\rm{p*}}} - I_{\rm{d}}^{\rm{p}}) + \omega LI_{\rm{q}}^{\rm{p}}},\\
{U_{\rm{q}}^{\rm{p}} = V_{\rm{q}}^{\rm{p}} - ({k_{\rm{p}}} + \frac{{{k_{\rm{i}}}}}{s})(I_{\rm{q}}^{{\rm{p*}}} - I_{\rm{q}}^{\rm{p}}) - \omega LI_{\rm{d}}^{\rm{p}}},
\end{array}} \right.
\end{align}
and similarly the negative-sequence node voltage is
\begin{align}
\left\{ {\begin{array}{*{20}{c}}
{U_{\rm{d}}^{\rm{n}} = V_{\rm{d}}^{\rm{n}} - ({k_{\rm{p}}} + \frac{{{k_{\rm{i}}}}}{s})(I_{\rm{d}}^{{\rm{n*}}} - I_{\rm{d}}^{\rm{n}}) - \omega LI_{\rm{q}}^{\rm{n}}},\\
{U_{\rm{q}}^{\rm{n}} = V_{\rm{q}}^{\rm{n}} - ({k_{\rm{p}}} + \frac{{{k_{\rm{i}}}}}{s})(I_{\rm{q}}^{{\rm{n*}}} - I_{\rm{q}}^{\rm{n}}) + \omega LI_{\rm{d}}^{\rm{n}}},
\end{array}} \right.
\end{align}
where ${U}$ denotes the node voltage of the converter, ${V}$ the grid voltage, ${I}$ the input voltage, ${L}$ the filtering inductance, ${\omega}$ the grid angular frequency, ${I^{*}}$ the reference input voltage, and ${k_{\rm{p}}} + \frac{{{k_{\rm{i}}}}}{s}$ the PI controller. The subscript (d vs.\ q) represents the respective components, and the superscript (p vs.\ n) differentiates the positive from the negative sequence.
\begin{figure}[h]
    \centering
\includegraphics[width=0.9\linewidth]{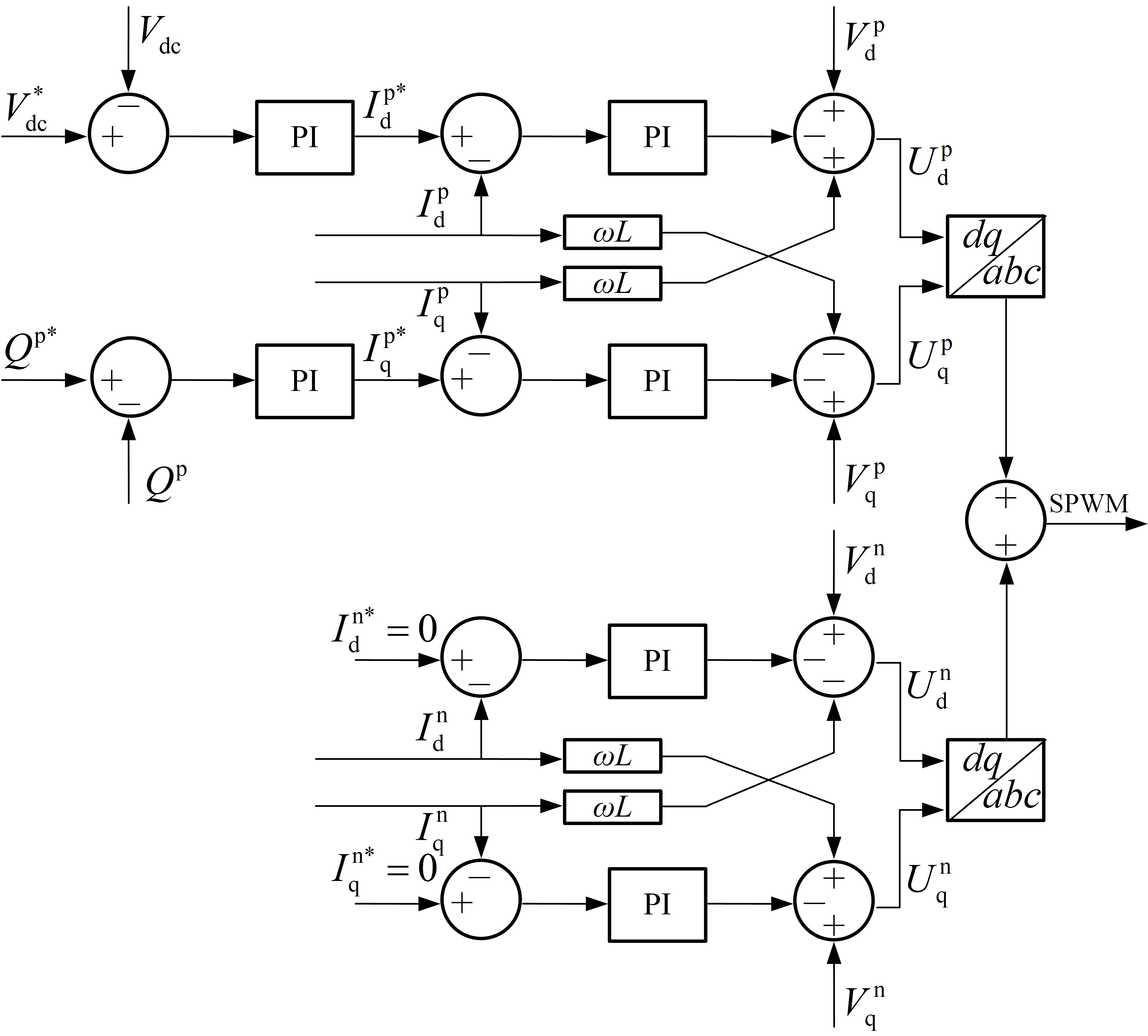}
    \caption{Shunt control block diagram.}
    \label{fig:PosNegDQ}
\end{figure}

In the control block diagram (Fig.~\ref{fig:PosNegDQ}), the dc-link voltage ${V_{\rm{dc}}}$ is regulated to the reference $V_{\rm{dc}}^{{\rm{*}}}$ for controlling the active current. The reactive component, which can inject currents to compensate the power factor of the grid, has a separate reference $Q^{{\rm{*}}}$. Both the active and reactive components of the negative sequence would be controlled to zero to not drive imbalance and can be used as a degree of freedom. For a balanced grid, it is sufficient to control only the positive sequence.

}

\subsection{Floating-Module Control}
\begin{figure}[h]
    \centering
    \includegraphics[width=0.45\linewidth]{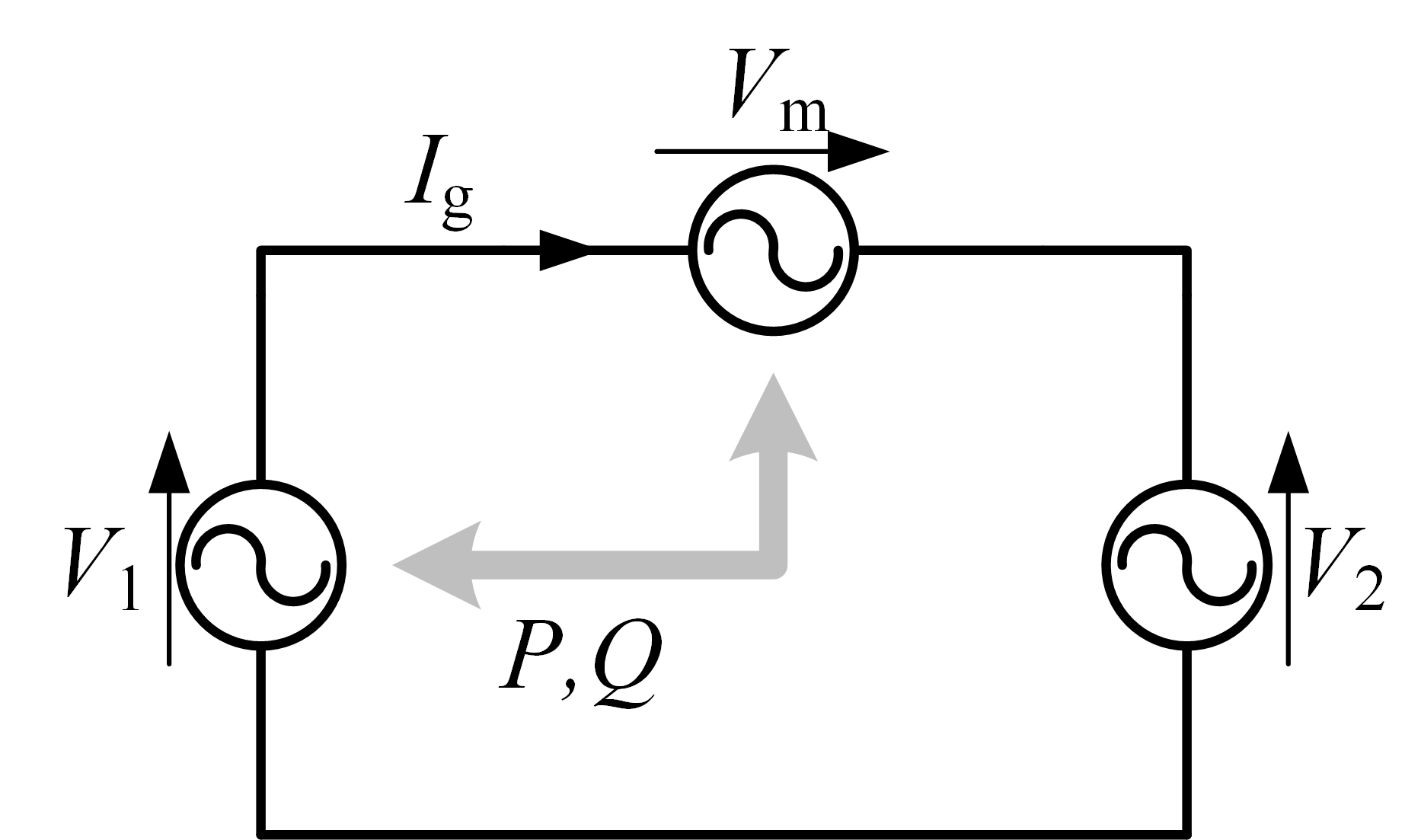}
    \caption{Equivalent circuit diagram of the floating module.}
    \label{fig:FM}
\end{figure}

As the key innovation of the solutions, the floating module performs direct electronic series injection. {As a voltage-sourced element, it injects a voltage difference between the left and right side of each phase (Fig.~\ref{fig:FM}).} An additional current control loop can regulate the flow. Consequently, the module may also be regarded as a controllable impedance. The dc-link voltage of each floating module is assumed to be relatively constant, which is provided by the AFE and translated to the module with a fixed ratio by the LLCs here.

If the floating module is inserted into the line in series, the line current can be derived as
\begin{align}
{I_{\rm{g}}} = \frac{{{{V_{\rm{2}}}} - {V_{\rm{1}}} - {V_{\rm{m}}}}}{{{Z_{{\rm{line}}}}}},
\end{align}
where ${{V_{\rm{2}}}}$ is the right terminal voltage, ${{V_{\rm{1}}}}$ is the left terminal voltage, ${{V_{\rm{m}}}}$ is the voltage actively generated by the floating module, and ${{Z_{\rm{line}}}}$ is the total line impedance.

Therefore, the power delivered/absorbed by the floating module follows
\begin{align}
{S_{\rm{m}}} = {V_{\rm{m}}} \cdot {I_{\rm{g}}} = {V_{\rm{m}}} \cdot \frac{{{V_{\rm{2}}} - {V_{\rm{1}}} - {V_{\rm{m}}}}}{{{Z_{{\rm{line}}}}}}.
\label{con:Sm}
\end{align}

In the two-grid or two-feeder case (assuming the right terminal has a higher potential than the left terminal), if the floating module injects exactly the difference between two grids, the power from the floating module will be zero; if the floating module injects only a bit more than the difference, it can reverse the power flow. In the mixed active-reactive load scenario, ${{V_{\rm{2}}}}$ is zero, and both active power and reactive power can be flexibly controlled according to Eq.~(\ref{con:Sm}).

\begin{figure}[h]
    \centering
    \includegraphics[width=0.9\linewidth]{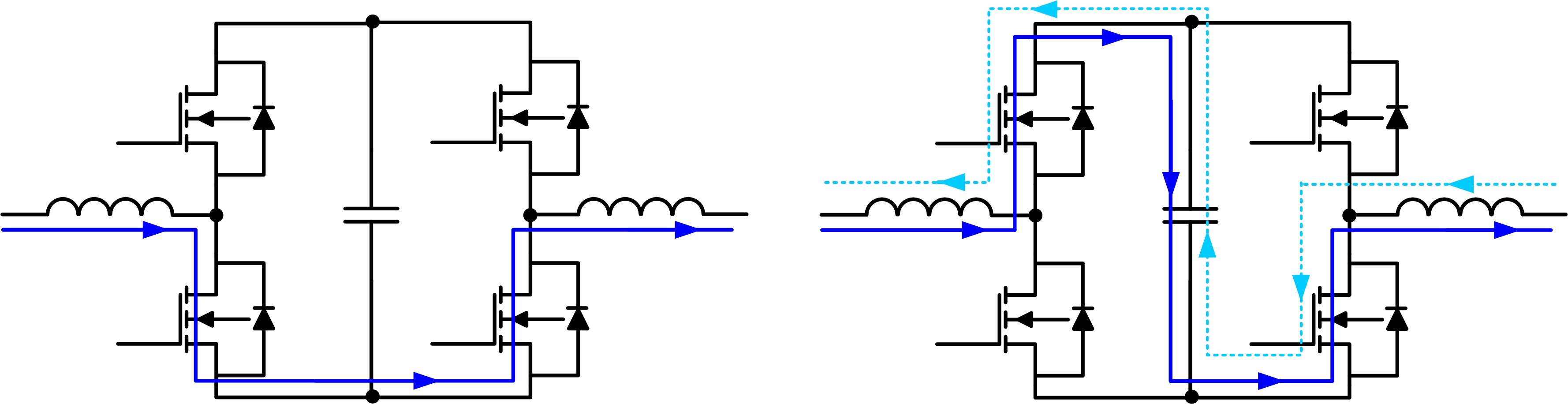}
    \caption{Working modes: bypass state (left); series state (right).}
    \label{fig:working_mode}
\end{figure}

Similar to cascaded bridge circuits, the series modules have at least three switching states, between which the modules can also modulate: { (1) a bypass state conducts the current around the floating dc-link capacitor and sets the voltage to zero; (2) two series states with positive and negative polarities temporarily connect the floating dc-link capacitor between the left and the right grid terminals, injecting a corresponding voltage (Fig.~\ref{fig:working_mode}).} A permanent bypass mode turns off flow control. In this case, the whole system is bypassed and the grid operates without conditioning. As latest low-voltage transistors can manage even low-voltage-grid short-circuit-level currents, this mode can further serve as a save state in case of failures.

Under flow control, each module injects an ac voltage with low amplitude across its terminals, which it modulates using the bypass and series modes. If the module connects, for instance two grid segments or feeders with different voltage and maintains exactly that voltage difference, the current flow is zero; with smaller or larger voltage, active current starts to flow, and the series module either extracts or injects energy from its dc link. Similarly, the module can compensate a phase shift between connected grid segments. For the fundamental frequency, the key degrees of freedom of this voltage difference are accordingly phase angle relative to the grid frequency and amplitude.

\begin{figure}[h]
    \centering
    \includegraphics[width=0.9\linewidth]{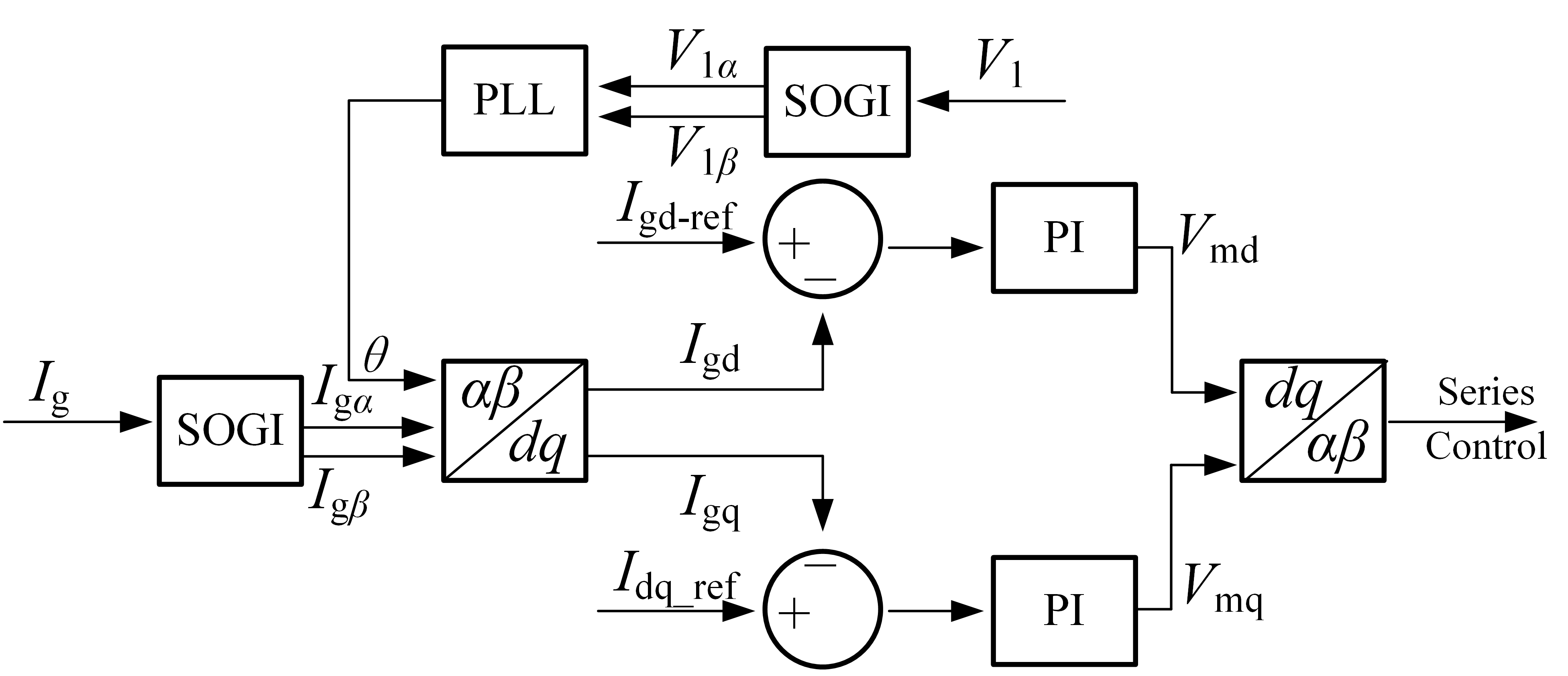}
    \caption{Series injection control block diagram.}
    \label{fig:Series_control}
\end{figure}

{ We suggest individual control of each module to allow for phase imbalance. We applied second-order generalized integrators (SOGI) for both single-phase grid synchronization as well as control and generate an orthogonal system for each phase \cite{single_phase_PLL}.} The generated SOGI signals and the single-phase model of each floating module can then be converted to d and q components as is established \cite{Multivariable_dq}. Fig.~\ref{fig:Series_control} shows a grid-synchronous current control scheme for the fundamental frequency. Each module has a dedicated control to handle phase asymmetries. For harmonics, the structure is repeated at the specific frequencies. With current–voltage control of the floating modules, open-loop operation of the LLCs, and the voltage–current control of the AFE, the power flow between AFE–LLC dc link and floating-module dc link is a consequential by-product. 

\section{Coverage Range Analysis}

The voltage of the floating modules performing the series injection determines the operation range of the concept. The following will analyze the possible compensation range of the power factor and amplitude differences within a grid mesh or two connected grid segments. Harmonics, in contrast, typically have relatively small voltage amplitudes compared to the fundamental. In the first approximation, their amplitude can be subtracted from the usable module voltage to identify the available effective voltage for flow control. The following discusses the two most prominent cases of a mixed active–reactive load connected to a grid and a grid mesh or two grid segments jointed through the series modules.
    
\subsection{Mixed active–reactive load represented by impedance (RL)}

\begin{figure}[h]
    \centering
    \includegraphics[width=0.8\linewidth]{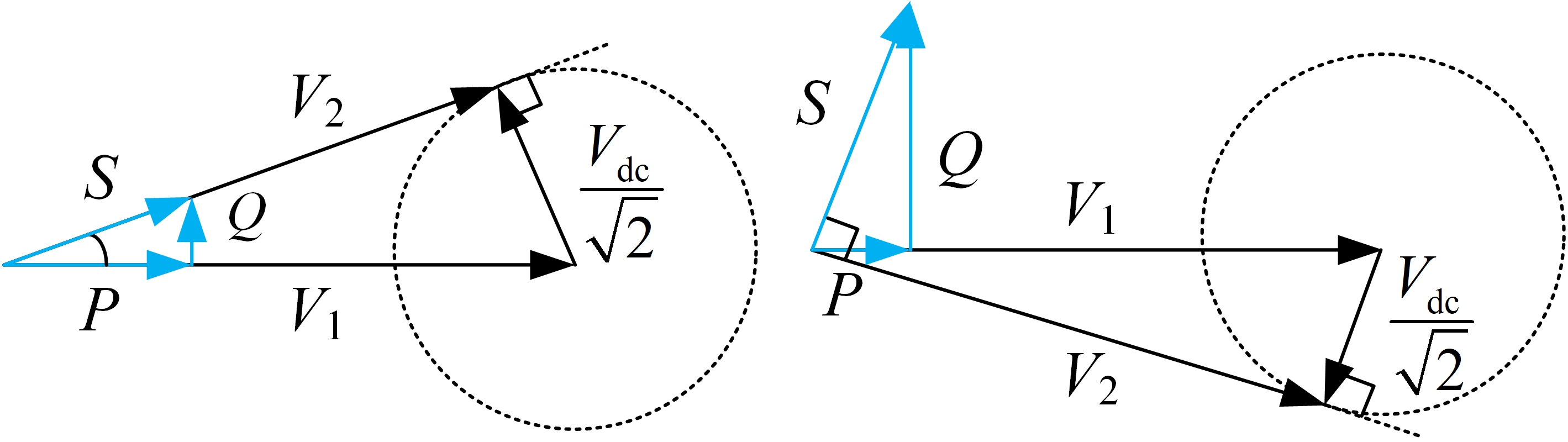}
    \caption{{Full reactive power compensation (left); full active power compensation (right).}}
    \label{fig:Analysis_RLload_active}
\end{figure}

The direct-injection power flow and quality controller can fully compensate and shield off the reactive power from a load {or the grid}. Fig.~\ref{fig:Analysis_RLload_active} shows the floating module voltage vector ${V_{{\rm{dc}}}}/\sqrt 2 $ (maximum amplitude without over-modulation) can rotate around the grid voltage vector ${V_{{\rm{1}}}}$, producing a resultant vector ${V_{{\rm{2}}}}$. {The interior of the circle represents the range of the floating module for adjusting the voltage amplitude or shifting its phase.} Therefore, without over-modulation, the reactive power and active power should fulfill the condition
\begin{align}
{\tan^{ - 1}}\left( {\frac{Q}{P}} \right) \le {\sin ^{ - 1}}\left( {\frac{{{V_{{\rm{dc}}}}}}{{\sqrt 2 {V_{\rm{1}}}}}} \right).
\label{con:full_reactive_comp}
\end{align}

Similarly, to enable full active power compensation, the reactive power and active power of the load should meet the condition
\begin{align}
{\tan ^{ - 1}}\left( {\frac{Q}{P}} \right) + {\sin ^{ - 1}}\left( {\frac{{{V_{{\rm{dc}}}}}}{{\sqrt 2 {V_{\rm{1}}}}}} \right) \ge \frac{\pi }{2}.
\label{con:full_active_comp}
\end{align}

Fig.~\ref{fig:Analysis_RLload_active} visualizes the two conditions. A dashed circle indicates the maximum voltage difference a floating module can generate as determined by its dc-link voltage ${V_{\rm{dc}}}$. Maximum in-phase or out-of-phase voltage injection ends at that circle. 

\begin{figure}[h]
    \centering
    \includegraphics[width=0.6\linewidth]{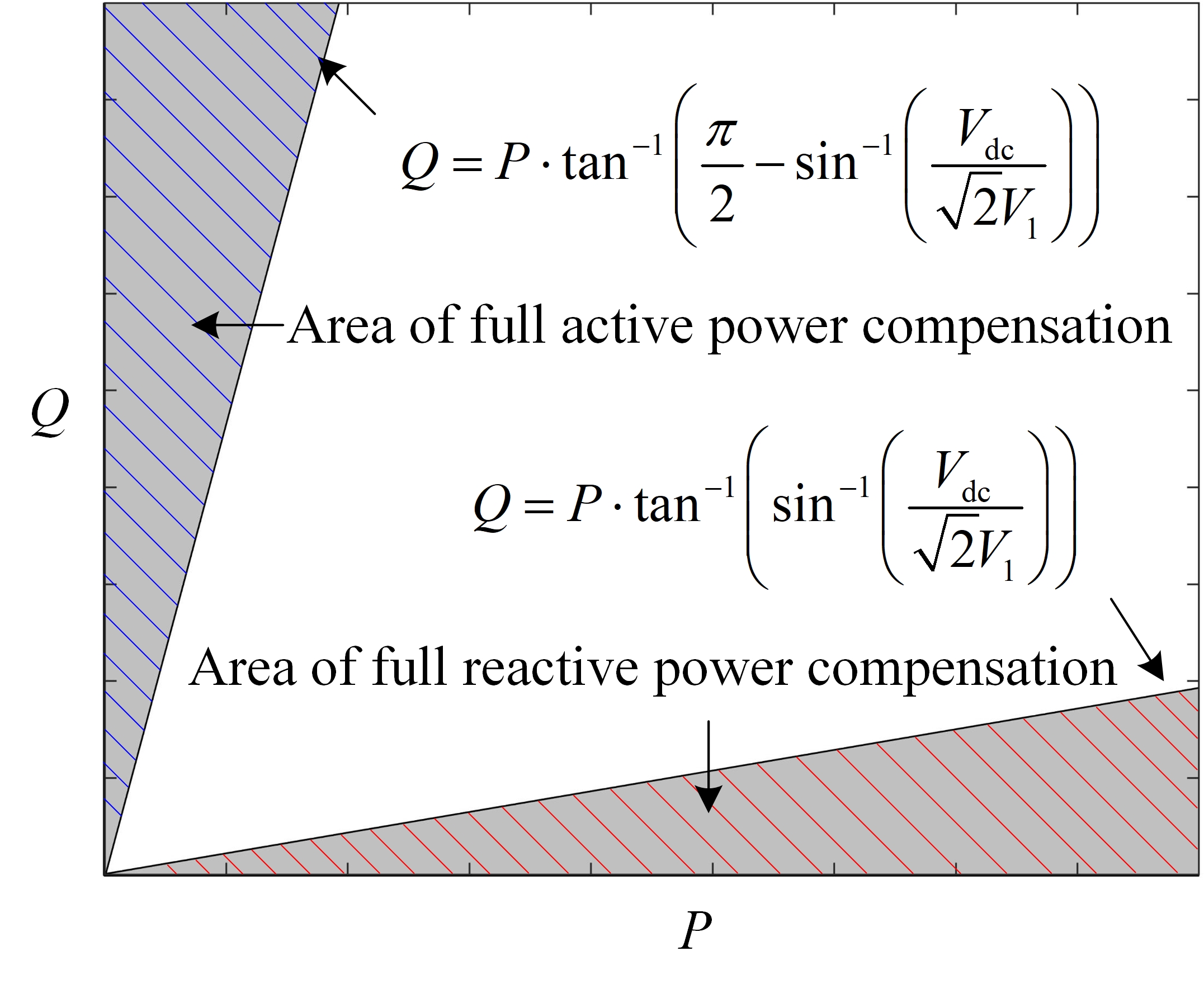}
    \caption{Coverage of the floating module in full active/reactive power compensation.}
    \label{fig:Coverage_active/reactive power_compensation}
\end{figure}

This limit translates into the hatched areas in the active–reactive power plane in Fig.~\ref{fig:Coverage_active/reactive power_compensation}. In the fully active/reactive power-compensation scenario, the acceptable ratio of active (P) and reactive power (Q) corresponds to the shifted circles constrained by the ratio of floating module voltage and the grid voltage.

\subsection{Two grids or feeders}
\begin{figure}[h]
    \centering
    \includegraphics[width=0.45\linewidth]{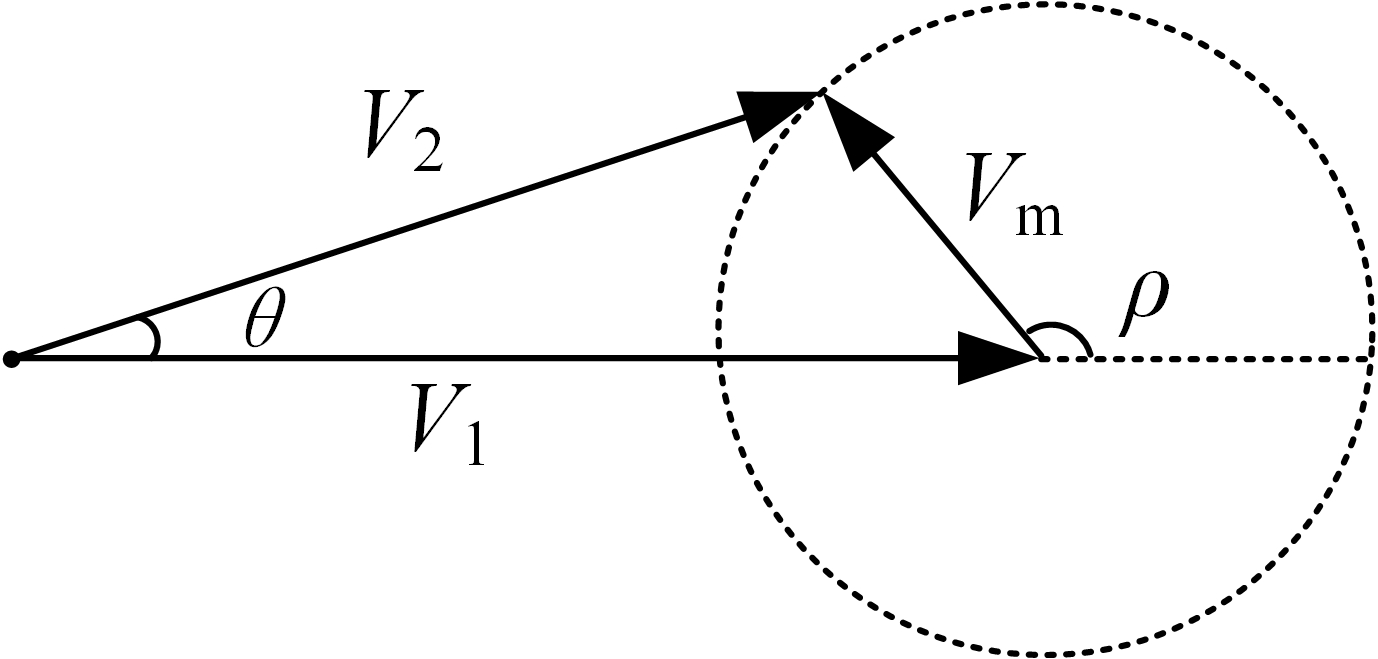}
    \caption{Phasor diagram of two grids or feeders.}
    \label{fig:oparea}
\end{figure}

{ When the floating modules are connected to grid segments or branches of a grid mesh on both sides, there can be differences in both voltage amplitude and phase between the grids. The phasor diagram in Fig.~\ref{fig:oparea} illustrates that the floating module can serve as a well controlled bridge to either stop current flow completely or enable a highly regulated flow between the two grid terminals. The variable names follow the equivalent circuit of this condition in Fig.~\ref{fig:FM}.}

To compensate for the voltage difference between two grids connected through the floating modules, the modules generate precisely the amount of voltage difference required to eliminate any current flow. However, if the voltage compensation is insufficient or shifted in phase, a residual current flows from the slightly higher-voltage side to the lower-voltage side. Depending on whether voltage or current control is used, the resulting current is determined by the grid impedances or can be directly regulated. The floating modules also absorb the power of the voltage difference multiplied by the currents flowing across them in their low-voltage dc links. This energy is transferred back to the high-voltage dc link by the LLCs. Additionally, the floating modules can reverse the direction of power flow by figuratively pumping active (and/or reactive) current against the voltage potential. This is achieved by setting a larger (and/or phase-shifted) voltage between both sides than the nominal voltage difference between them. 

As a result, the voltage of the dc link of the floating modules limits the maximum amplitude and phase difference that can exist between two grid segments or grids whilst still allowing the power-flow controller to connect them. The amplitude difference $\Delta V$ and the phase difference $\Delta \theta$ follow
\begin{equation}
\Delta V = \left| {{V_{{\rm{2}}}} - \\ \left(\sqrt {{V_{\rm{m}}}^2 - {{({V_{{\rm{2}}}} \cdot \sin \Delta \theta )}^2}} + {V_{{\rm{2}}}}\cos \Delta \theta \right)} \right|,
\end{equation}
where ${V_{{\rm{2}}}}$ is the voltage of one grid, { ${V_{{\rm{1}}}}$ of the other one, and ${V_{{\rm{m}}}}$ the differential output voltage of the floating module.}

\begin{figure}[h]
    \centering
    \includegraphics[width=0.8\linewidth]{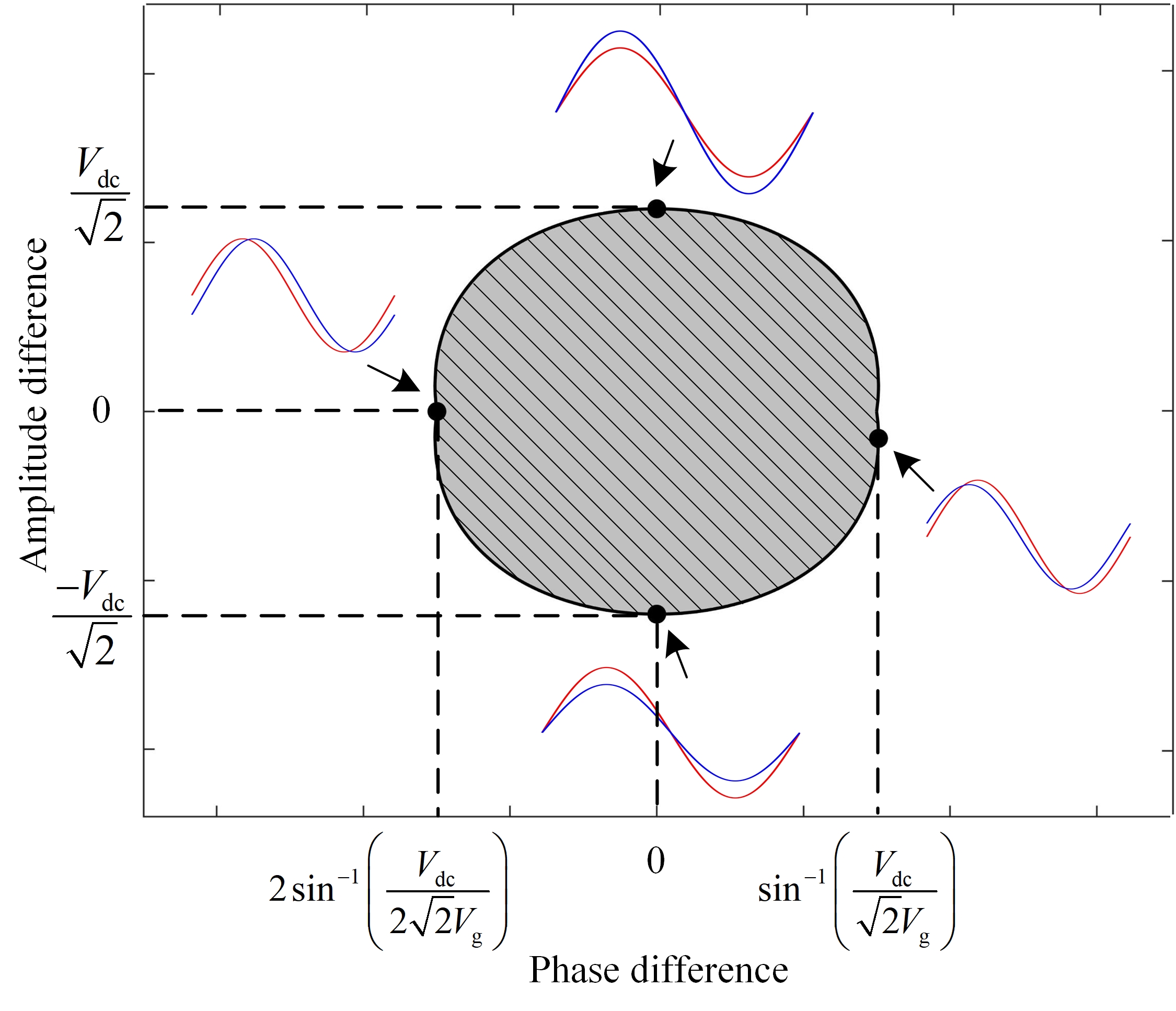}
    \caption{Coverage of the floating module connected between two grids or feeders.}
    \label{fig:magnify}
\end{figure}

Fig.~\ref{fig:magnify} illustrates the operation area. The actual grid voltage ${V_{\rm{g}}}$ is defined as the average of two grids' voltage (i.e., ${V_{\rm{g}}} = \left( {{V_{ {{\rm{2}}} }} + {V_{ {{\rm{1}}}}}} \right)/2$) to avoid defining a nominal side so that the system and the plot are symmetric. With over-modulation, the range can be extended to the outer limit \cite{over_modu}. However, over-modulation would necessarily inject harmonics and therefore rather serves as a fall-back option in critical situations.

The figure marks the four outermost points in terms of voltage and phase direction, representing different grid voltage conditions. The first point on the left represents the case where the voltage of both grids is the same but with a maximum phase shift. The second point on the right represents the case where there is the maximum global phase difference between the two grids. The third and fourth points represent the case where there is a maximum in-phase voltage difference, with one point being above and the other below. The sine wave plots illustrate the different voltage conditions. For a typical case where the nominal effective grid voltage is 230~V and the floating-module voltage is 48~V, the maximum limits for amplitude and phase difference would be 34~V effective and 8.49° (48~V and 12° with over-modulation), respectively. { The 48 V level is well established and cultivated by the automotive industry (defined as a band per industry standard LV148 there) so that a highly cost-optimized environment including semiconductors exists. That level is also perfect for bridging the maximum tolerance band of the low-voltage grid of ±10\% per IEC 60038. Thus, developments in other areas, such as automotive, in combination with the coincidental almost ideal fit with the tolerance levels in the grid enabled this new circuit.}

\begin{figure}[h]
    \centering
    \includegraphics[width=0.8\linewidth]{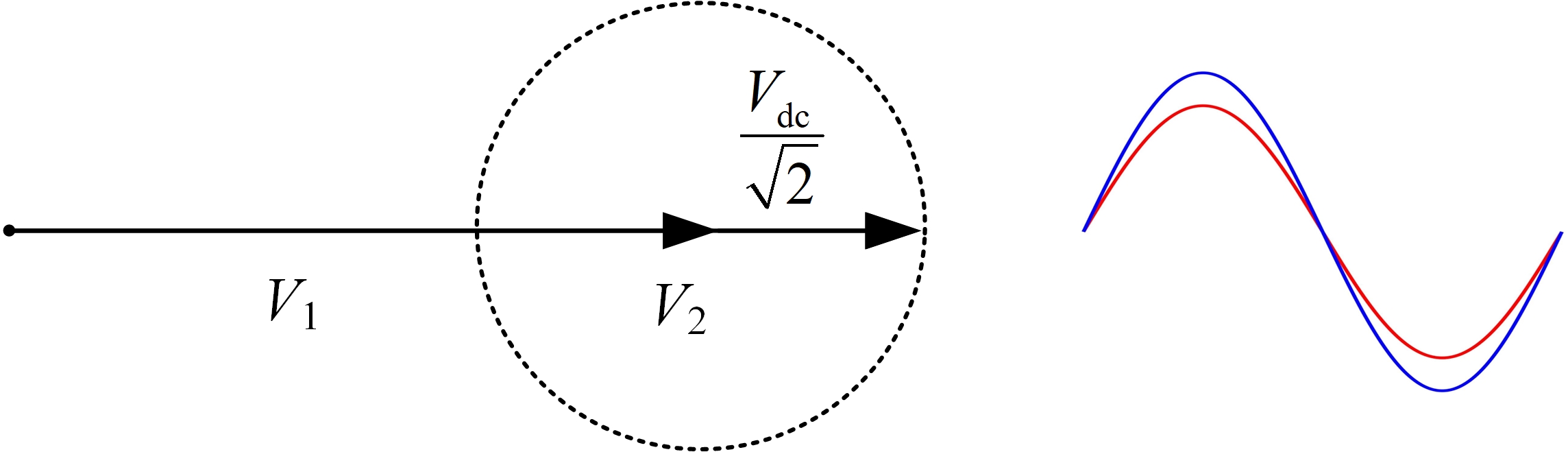}
    \caption{Phasor diagram of two in-phase grids with maximum voltage difference without over-modulation.}
    \label{fig:AnalysisMaxVDiff}
\end{figure}

Fig.~\ref{fig:AnalysisMaxVDiff} illustrates the case of the maximum in-phase voltage difference, where the floating modules operate at the {maximum} modulation index without over-modulation, which leads to $\left| {{V_{ 2 }} - {V_{1}}} \right| \le {V_{{\rm{dc}}}}/\sqrt 2 $ ($\left| {{V_{2}} - {V_{1}}} \right| \le {V_{{\rm{dc}}}}$ with over-modulation). 

\begin{figure}[h]
    \centering
    \includegraphics[width=0.8\linewidth]{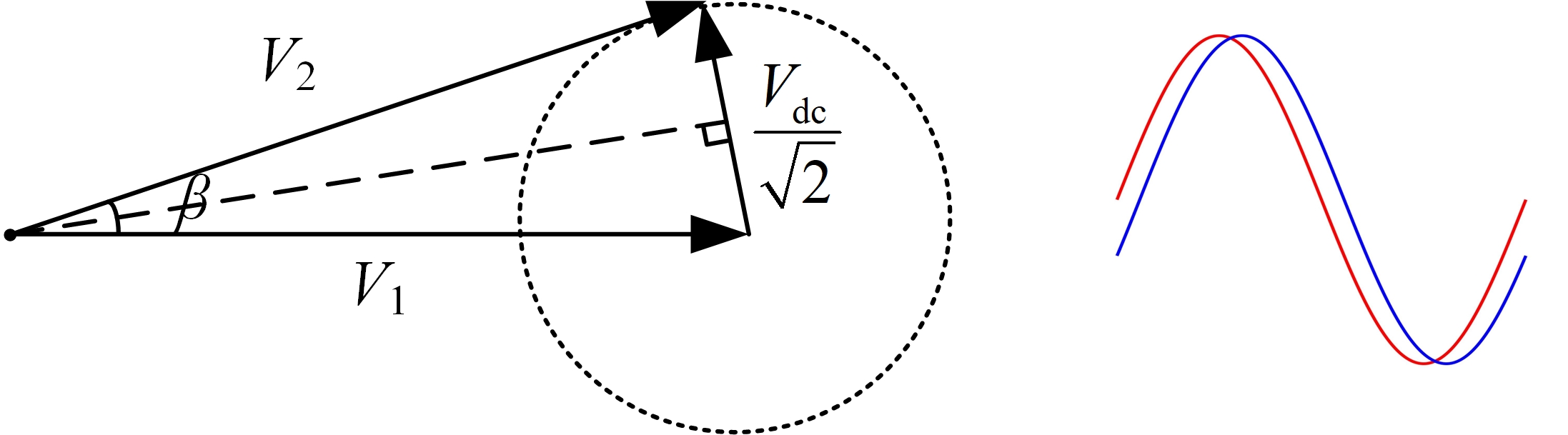}
    \caption{Phasor diagram of two equal-amplitude grids with maximum phase difference without over-modulation.}
    \label{fig:AnalysisMaxPhaseDiff}
\end{figure}

Fig.~\ref{fig:AnalysisMaxPhaseDiff} represents the case of voltage parity but maximum phase shift without over-modulation. The phase difference $\beta$ in this situation follows Eq.~(\ref{con:voltage_parity_maximum_phase_shift}).
\begin{equation}
{V_{\rm{1}}} \cdot \sin\left( {\frac{\beta }{2}} \right) = \frac{{{V_{{\rm{dc}}}}}}{{2\sqrt 2 }}
\label{con:voltage_parity_maximum_phase_shift}.
\end{equation}

\begin{figure}[h]
    \centering
    \includegraphics[width=0.8\linewidth]{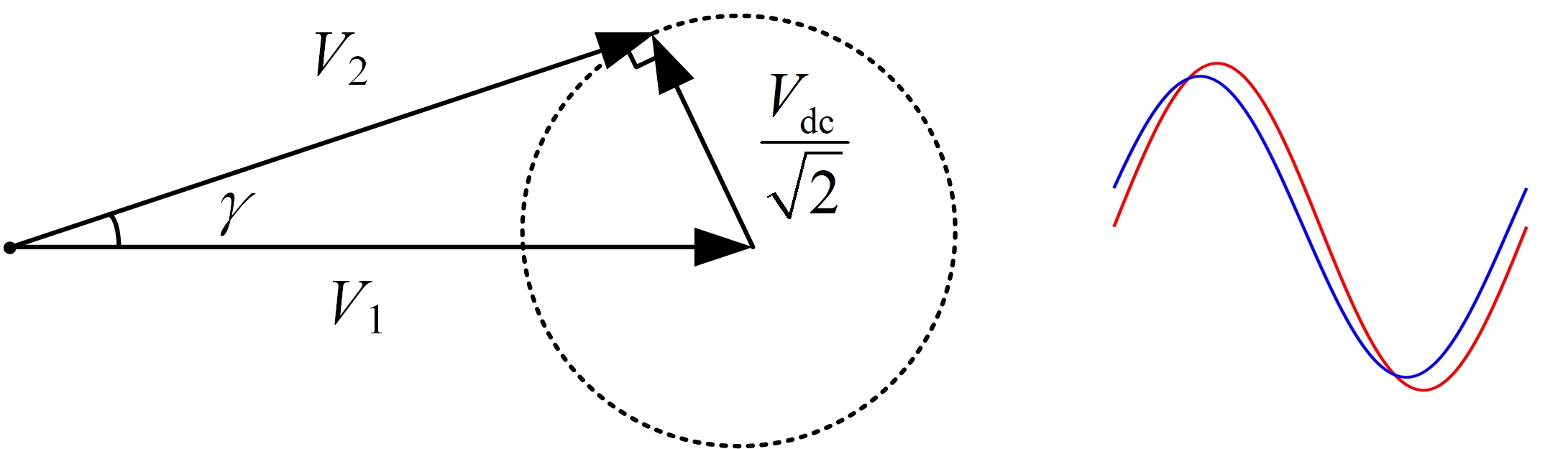}
    \caption{Phasor diagram of two grids with global phase difference without over-modulation.}
    \label{fig:AnalysisMaxPhaseDiffGlobal}
\end{figure}

The global phase difference maximum occurs when one grid’s phasor is tangent to the circle floating module’s phasor forms, as shown in Fig.~\ref{fig:AnalysisMaxPhaseDiffGlobal}. The voltages of the grids connected on the left and right sides necessarily differ in that case. The global phase difference maximum $\gamma$  follows
\begin{equation}
{V_{\rm{1}}} \cdot \sin\left( \gamma  \right) = \frac{{{V_{{\rm{dc}}}}}}{{\sqrt 2 }}
\label{con:AnalysisMaxPhaseDiffGlobal}.
\end{equation}

As for the active/reactive regulation range, assuming the line impedance is $j{X_{{\rm{line}}}}$, the apparent power delivered from one feeder without control follows
\begin{equation}
\begin{split}
{S_{{\rm{1\to2}}}} &= {P_{{\rm{1\to2}}}} + j{Q_{{\rm{1\to2}}}} \\
&= - \frac{{{V_{{\rm{1}}}}{V_{{\rm{2}}}}\sin \theta }}{{{X_{{\rm{line}}}}}} + j\frac{{{V_{{\rm{1}}}}^2 - {V_{{\rm{1}}}}{V_{{\rm{2}}}}\cos \theta }}{{{X_{{\rm{line}}}}}},
\end{split}
\end{equation}
where ${P_{{\rm{1\to2}}}}$ is the active power delivered by Grid Segment 1, ${Q_{{\rm{1\to2}}}}$ the reactive power delivered by Grid Segment 1, and $\theta $ the angle between the two grid segments or grids (see Fig.\ \ref{fig:oparea}).

The relationship between active and reactive power can be expressed as
\begin{equation}
{P_{{\rm{1\to2}}}}^2 + {\left({Q_{{\rm{1\to2}}}} - \frac{{{V_{{\rm{1}}}}^2}}{{{X_{{\rm{line}}}}}}\right)^2} = {\left(\frac{{{V_{{\rm{1}}}}{V_{{\rm{2}}}}}}{{{X_{{\rm{line}}}}}}\right)^2}.
\end{equation}
When the floating module modulates, the apparent power changes to
\begin{equation}
{S_{{\rm{1\to2}}}}' = {S_{{\rm{1\to2}}}} - \frac{{{V_{{\rm{1}}}}{V_{\rm{m}}}\sin \rho }}{{{X_{{\rm{line}}}}}} - j\frac{{{V_{{\rm{1}}}}{V_{\rm{m}}}\cos \rho }}{{{X_{{\rm{line}}}}}},
\end{equation}
where $\rho$ is the phase angle of the floating module ($\rho \in [0,2\pi)$), and ${V_{{\rm{m}}}}$ is the output voltage generated by the floating module, reaching its maximum at ${{V_{{\rm{dc}}}}/\sqrt 2}$.

\begin{figure}[h]
    \centering
    \includegraphics[scale = 0.8]{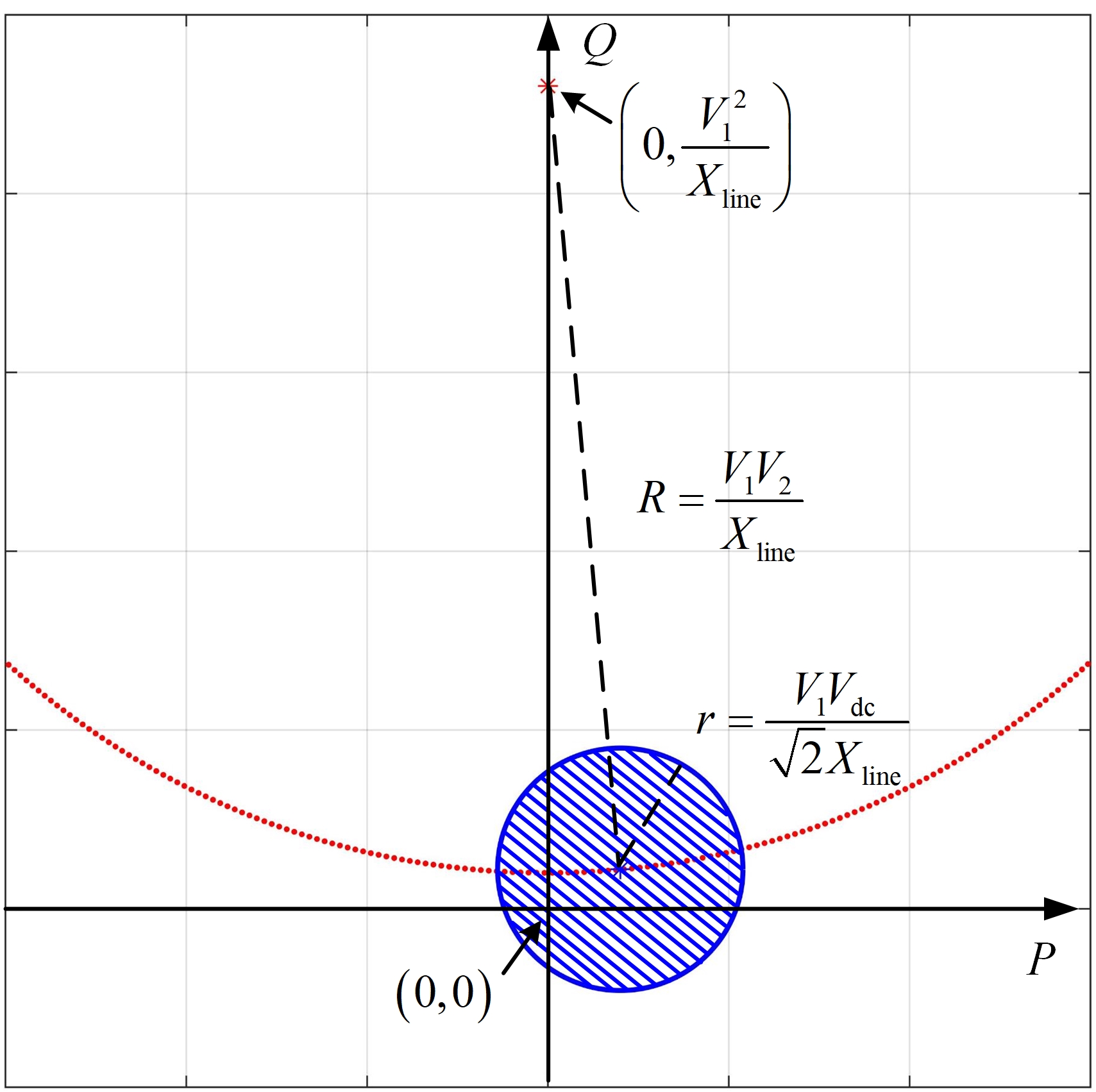}
    \caption{Power regulation coverage of the floating module between two grids or feeders.}
    \label{fig:regulation3}
\end{figure}

Therefore, the control region can be seen in Fig.\ \ref{fig:regulation3}. The red curve represents the active and reactive power delivered by Grid Segment 1 with varying phase angle $\theta $. The floating module can move along the red curve and cover a blue circular range with a radius varying with 
${V_{{\rm{1}}}}{V_{{\rm{m}}}}/ {X_{{\rm{line}}}}$, maximizing at ${V_{{\rm{1}}}}{V_{{\rm{dc}}}}/\sqrt2 {X_{{\rm{line}}}}$. The center of the circle is determined by the phase angle $\theta $. As the circle covers four quadrants, the floating module can freely regulate both active and reactive power. 

For a typical distribution grid, where the line impedance can reach the order of 0.1~$\Omega$ \cite{Lowvoltage_impedance}, let the two grid segments have 230~V and 220~V respectively, the maximum power that can in that example be adjusted by the floating module at the grid segment connected to the right side is consequently ±~78~kW~/ ±~78~kVar. For lower impedance values, the power levels increase correspondingly. 

{ The proposed circuit has the same compensation range as a conventional transformer-based UPFC/UPQC but excels in cost and size \cite{Understanding_FACTS}. Unlike a full back-to-back converter with semiconductors for the full grid power rating in voltage and current, the proposed circuit merely needs to withstand the voltage difference between two feeders and convert a small power to control approximately ten-fold higher power levels.}

\section{Simulations}
We modelled the circuit and associated control in Matlab/Simulink to analyze the performance. We set the proportional and integral gains of the floating module to 0.1 and 1 respectively (see Fig.~6). The line rating provides the necessary power-flow reserve. The grid and the floating-module voltage rating determine the required dc-link voltage as derived analytically above. { We analyzed six situations}: mixed active–reactive load (RL) with only active power passing through, mixed active–reactive load with only reactive power passing through, two grids with different amplitude, two grids with different phases, mixed active–reactive load with harmonics, { and an unbalanced grid}. The relative parameters are listed in TABLE I.
\subsection{Mixed active–reactive load with active power only}
\begin{figure}
    \centering
    \includegraphics[width=0.9\linewidth]{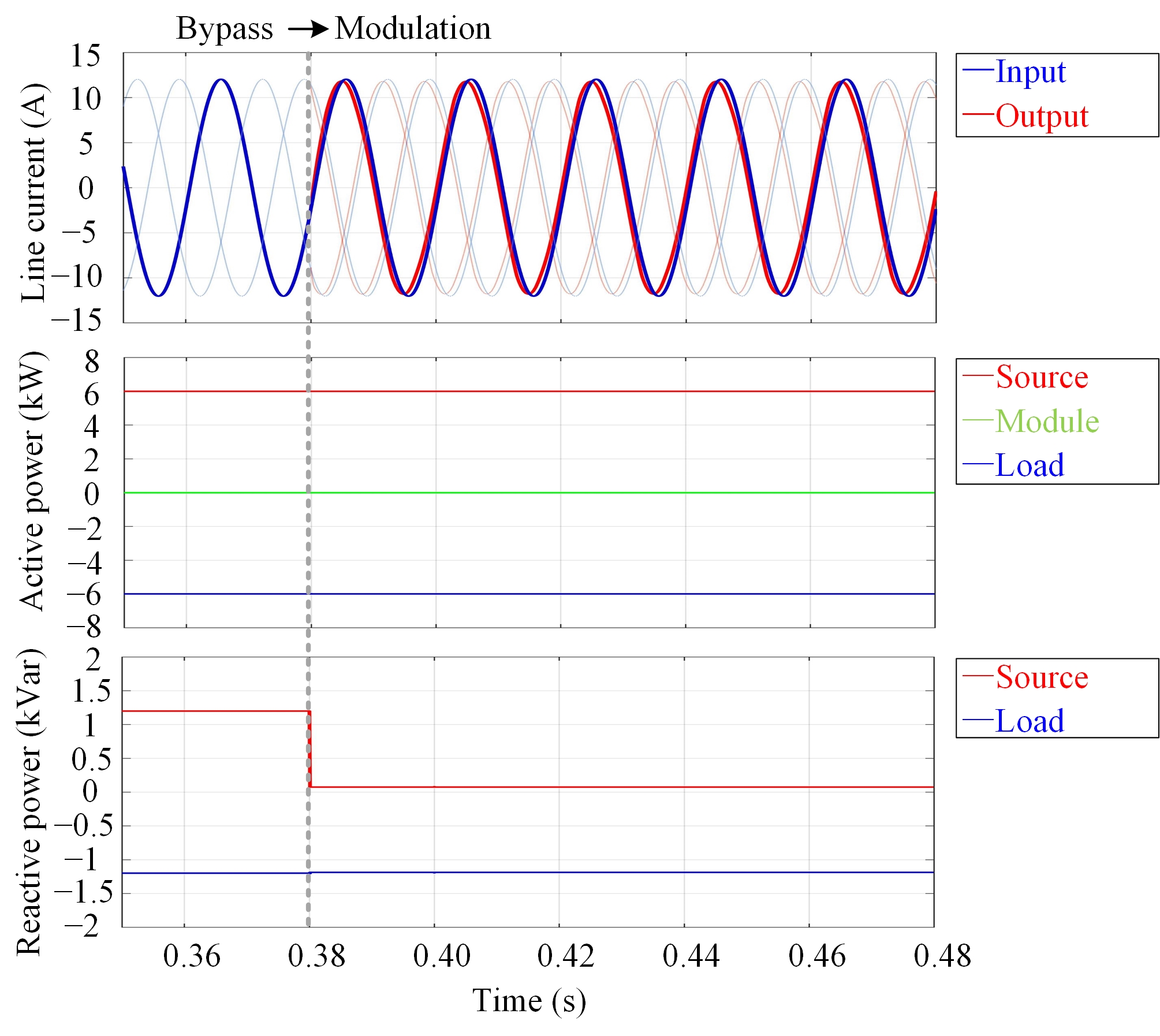}
    \caption{Current, active power, and reactive power for the RL load case (only active power passing through).}
    \label{fig:Onegridactive}
\end{figure}

\begin{table}
\centering
\caption{Test Parameters}
\centering
\begin{tabular}{l c }
\hline
Description & Value\\
\hline
Floating module parameters & 48~V, 100~kHz switching\\
Grid parameters & 400~V (ph--ph), 0\textdegree, 50~Hz\\
Line impedance per phase & (20 + j~10)~m$\Omega$ \\
Filter inductance per phase & 200~µH \\
\hline
{ Case A: Load impedance}  & {(22 + j~4)$~\Omega$} \\
{ Case B: Load impedance}  & {(4 + j~22)$~\Omega$} \\
{ Case C: Second grid parameters}  & { 390~V (ph--ph), 0\textdegree, 50~Hz}\\
{ Case D: Second grid parameters}  & { 400~V (ph--ph), 8\textdegree, 50~Hz}\\
{ Case E: Harmonics}  & { \makecell{3-th order with 0.1~pu and --25\textdegree\\ + 5-th order with 0.05~pu and 35\textdegree}} \\ 
{Case F: Phase imbalance}  &  {\makecell{Phases a, b, c: 245~V, 230~V, 200~V (rms)\\load: 5$~\Omega$}} \\
\hline
\end{tabular}
\end{table}
As previously analyzed, the floating module has the ability to compensate for all reactive power within the limits outlined in Eq.~(\ref{con:full_reactive_comp}). In Fig.~\ref{fig:Onegridactive}, the controller switches from bypass to modulation at t = 0.38~s, which causes a phase shift between the grid and load voltages. As a result, the module also shifts the line current forward in time (to the left) compared to the first part of the recording when there was no modulation (indicated by dark lines for the first phase and faint lines for the second and third phases). The input current (the floating module in bypass mode) is represented by the blue line, while the output current (the floating module in modulation mode) is represented by the red line. When the module is modulating, it shields the reactive power from the source grid, causing its contribution to drop from 1.2~kVar to approximately 0. The reactive power is maintained and delivered by the floating modules, where the dc-link capacitor absorbs the power fluctuation before it enters the grid.
\begin{figure}
    \centering
    \includegraphics[width=0.9\linewidth]{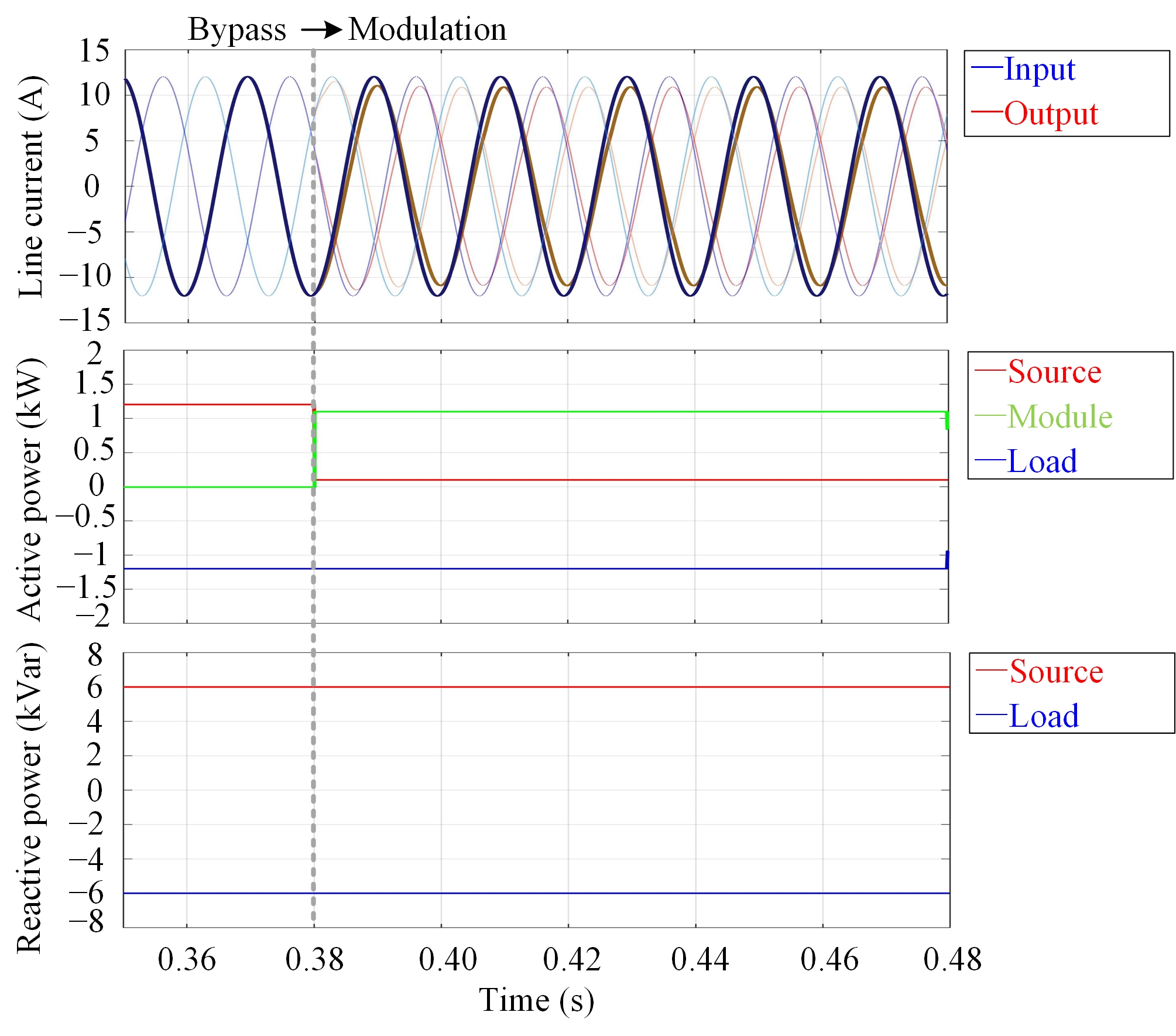}
    \caption{Current, active power, and reactive power for the RL load case (only reactive power passing through).}
    \label{fig:Onegridreactive}
\end{figure}

When viewed from the perspective of the line end, we can observe that the reactive components are removed, resulting in a phase difference between the line current in modulation mode (indicated by the red line) and its unchanged current (indicated by the blue line). The lighter lines represent the other two phases. As a result of this correction, the electrical system is now operating at the unity power factor, as viewed from the source end, and the grid voltage is perfectly in phase with the line current. Throughout the process, active power remains constant, as { the control approach does not aim to affect the active power.} This illustrates that the direct-injection f/q controller can effectively and accurately achieve a fully active power compensation mode.

\subsection{Mixed active–reactive load with reactive power only}

In a similar manner, when active and reactive power satisfy the conditions specified in Eq.~(\ref{con:full_active_comp}), the floating module can eliminate all active power and allow only reactive power to flow through the line to the grid. In this scenario, all active power is supplied through the dc link of the floating modules. In Fig.~\ref{fig:Onegridreactive}, the floating modules switch from bypass to active control at $t$ = 0.38 s and prevent active power from flowing from the grid through the series modules. Instead, the floating modules provide it through the LLCs and AFE to the load. As a result, the line current changes to purely reactive current and lags behind compared to the current in the bypass mode. 

\subsection{Two grids (same phase but different amplitude)}
\begin{figure}
    \centering
    \includegraphics[width=0.9\linewidth]{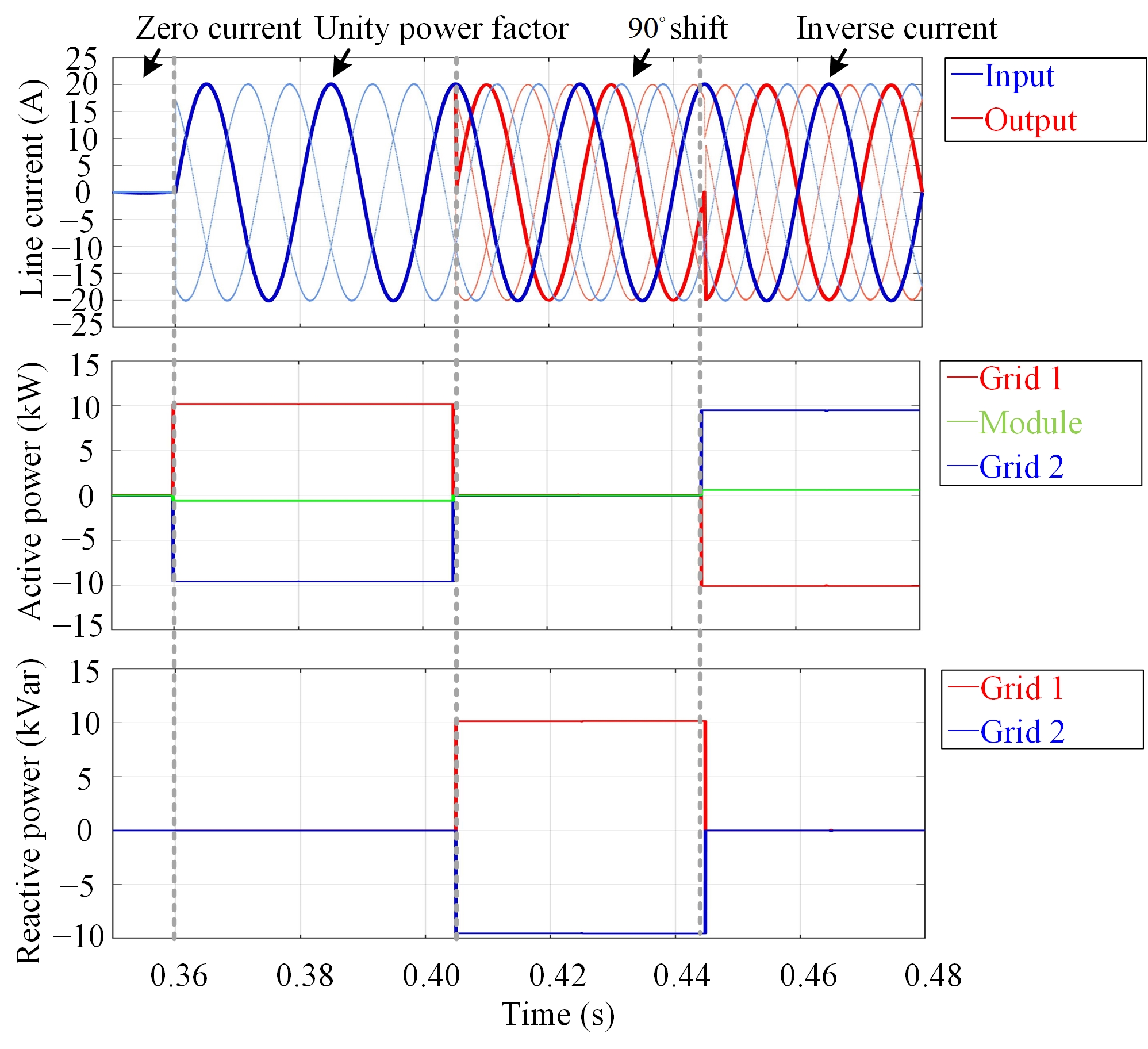}
    \caption{Current, active power, and reactive power for the two grids case (same phase and different amplitude).}
    \label{fig:amp}
\end{figure}

In Fig.~\ref{fig:amp}, the floating modules are inserted between two grids or grid segments with different voltage amplitude. If flow regulation is not implemented, a large current would flow into the lower-voltage segment and potentially damage the distribution cable. Therefore, the power-flow controller is designed to prevent any current flow by generating a series voltage initially. Starting from 0.36 seconds, the series modules control the current to a maximum of 20~A from the higher-voltage segment to the lower-voltage segment with unity power factor. The floating modules consume any excess power. In the following interval, the line current is shifted by 90°, resulting in a purely reactive power flow.
To demonstrate the full bidirectional capability, in the last interval, the current is shifted by another 90°, leading to active power flow from the lower-voltage to the higher-voltage grid segment against the voltage gradient. The dc link provides the necessary power to push the current against the voltage.
\begin{figure}
    \centering
    \includegraphics[width=0.9\linewidth]{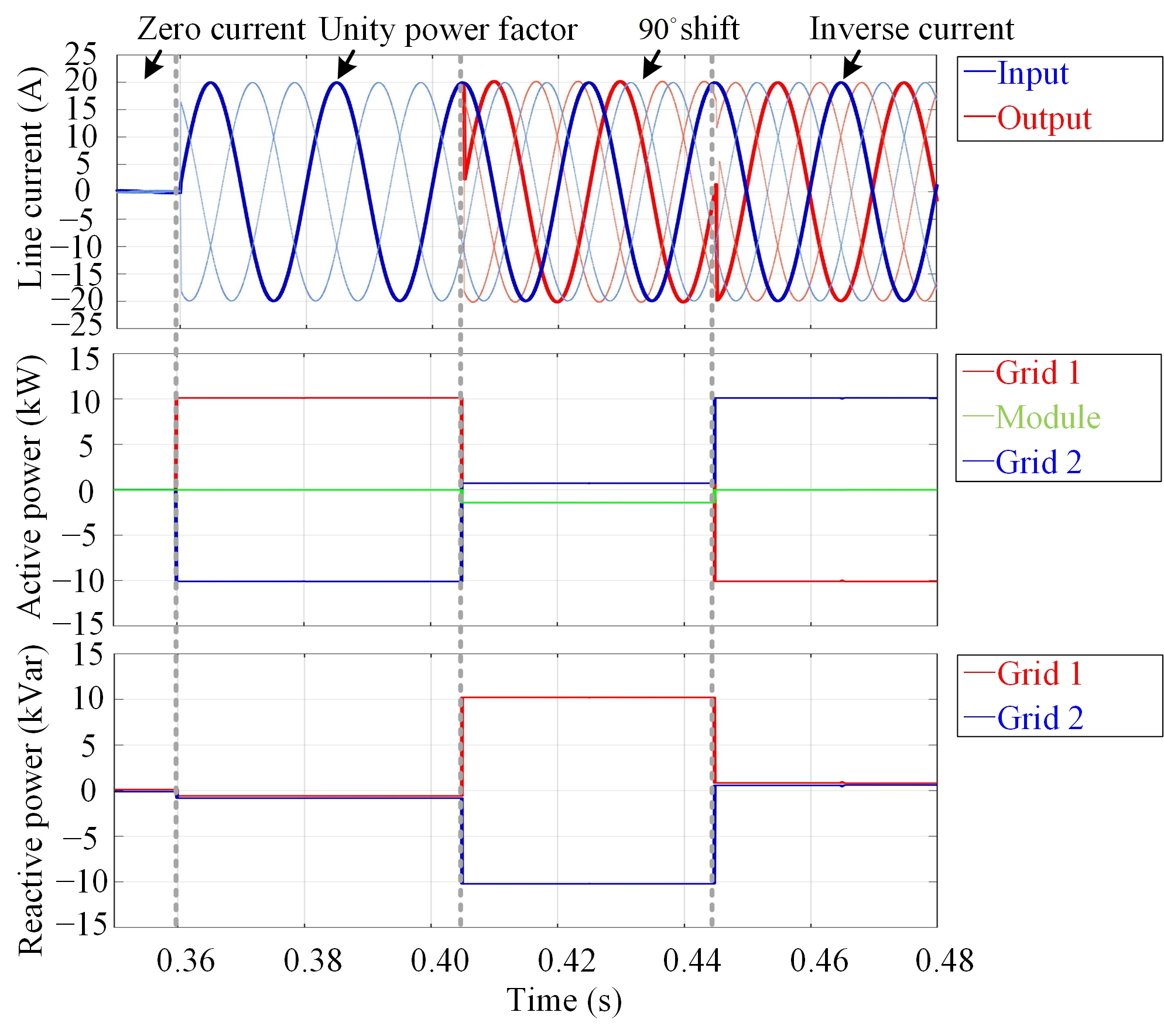}
    \caption{Current, active power, and reactive power for the two grids case (same amplitude and different phase).}
    \label{fig:phase}
\end{figure}

\subsection{Two grids (same amplitude but different phases)}
When using the power-flow controller to connect grid segments with the same voltage but different phase, the bypass mode would result in excessive current flow limited only by the grid impedance.  To prevent this, the controller maintains zero current flow in the initial interval shown in Fig.~\ref{fig:phase}, which creates the required voltage difference. The figure also illustrates the bidirectional flow of active and reactive power, which is comparable to the situation with grids having different voltage levels.

\subsection{{ Load with harmonic content}}
\begin{figure}
    \centering
    \includegraphics[width=0.9\linewidth]{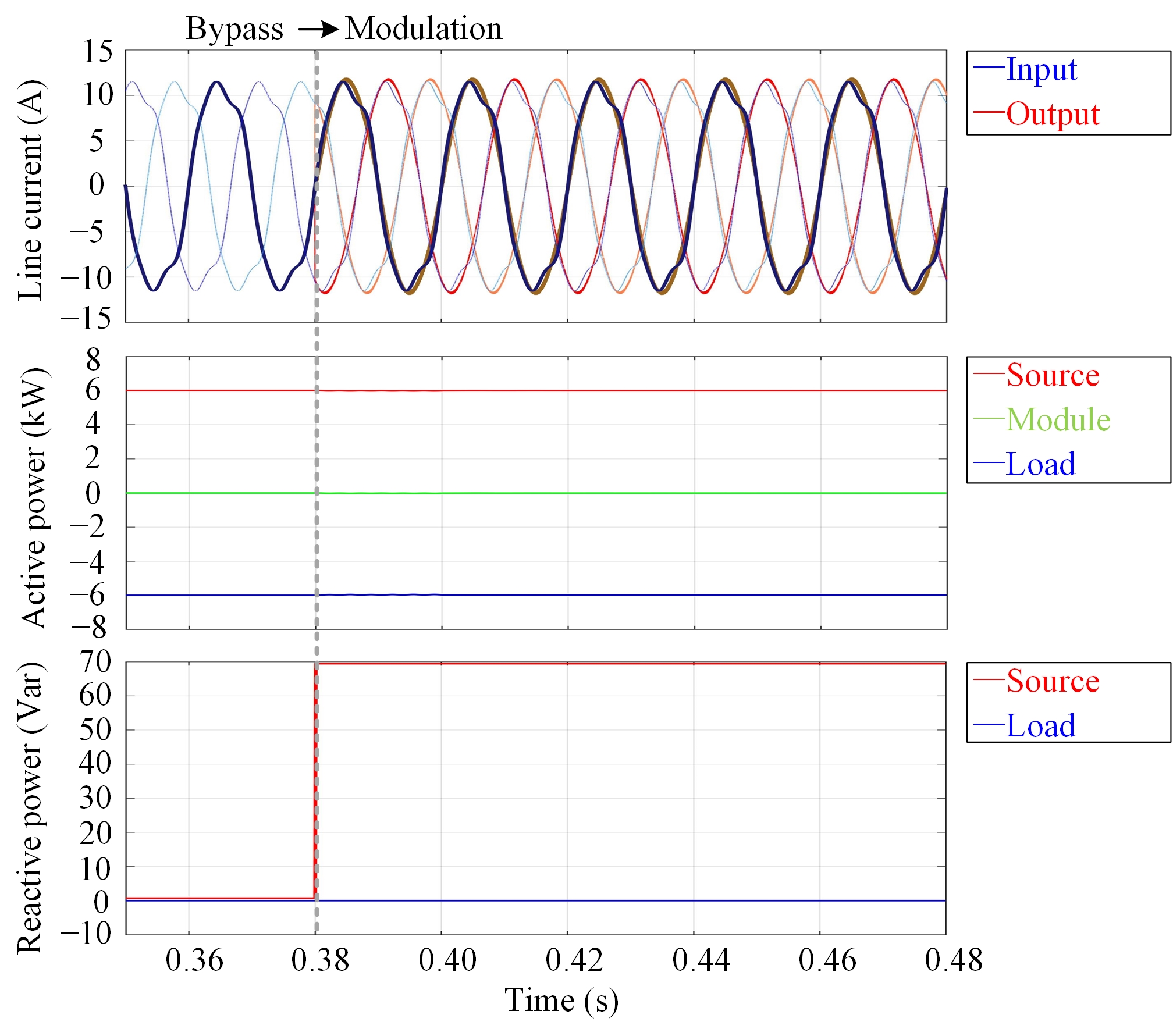}
    \caption{Current, active power, and reactive power for the RL load with harmonics.}
    \label{fig:harm}
\end{figure}

{ The series modules have the ability to filter out harmonics, i.e., block them, in addition to controlling the fundamental current. Fig.~\ref{fig:harm} illustrates a case where a 3rd harmonic (i.e., zero sequence) with 0.1~pu and a 5th harmonic with 0.05~pu pass through the power-flow controller until the controller suppresses them from $t$ = 0.38~s onward, allowing only the fundamental to pass through and produce sinusoidal conditions.} 
{ \subsection{Unbalanced grid}
Fig.~\ref{fig:unb} illustrates an instance where Phase a is 0.07~pu above and Phase c 0.13 pu below the nominal value. Each floating module, at its own phase, can work independently in phase with its phase's voltage, to address the unbalanced grid issue. Since this case uses resistive loads, the figure only displays the active power handled by the floating module.

\begin{figure}
    \centering
    \includegraphics[width=0.9\linewidth]{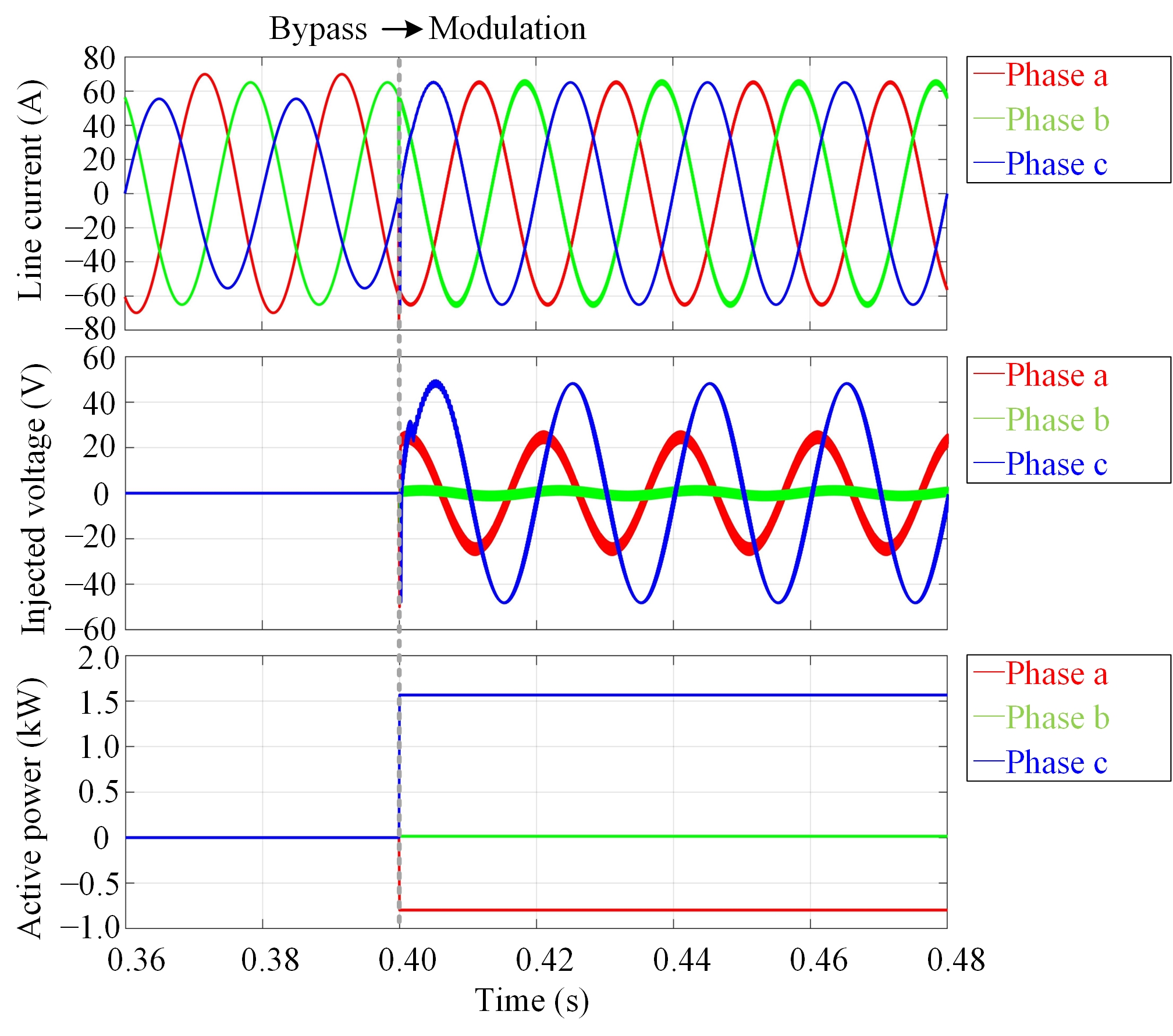}
    \caption{Current, injected voltage, and active power of the floating module for the unbalanced load.}
    \label{fig:unb}
\end{figure}
}

\section{Experimental Demonstration}
\subsection{Setup}
\begin{figure}
    \centering
    \includegraphics[width=0.9\linewidth]{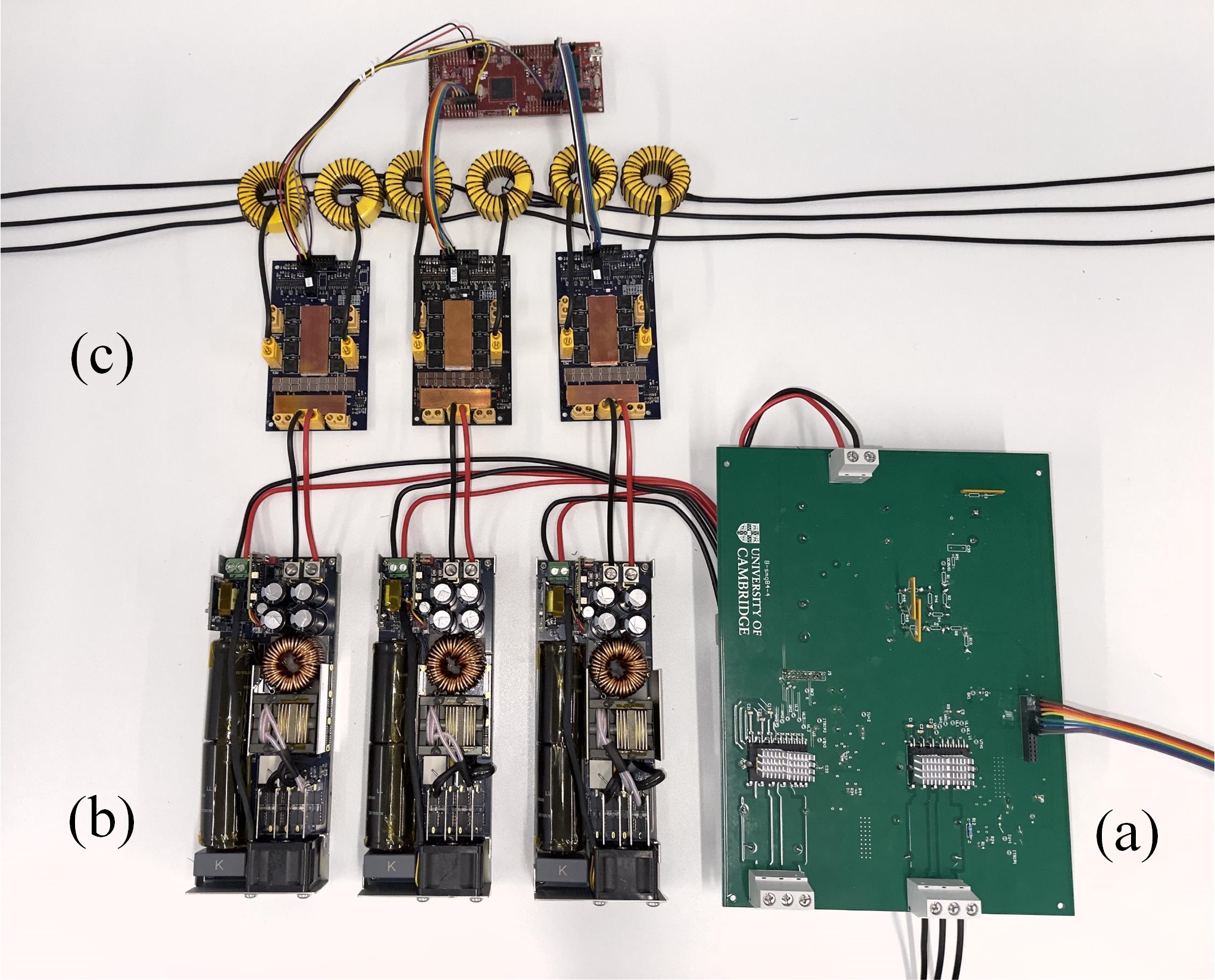}
    \caption{Experimental platform with (a) grid-voltage low-current shunt part, (b) dc/dc converter, and (c) low-voltage high-current floating series modules. The grid line runs as black cables from left to right. The setup is controlled by a DSP (uppermost circuit board).}
    \label{fig:setup_my}
\end{figure}

We implemented an experimental system to further substantiate the simulations, as shown in Fig.\ \ref{fig:setup_my}  ((a) an active front end; (b) three single-phase bidirectional dc/dc converters; (c) three single-phase floating modules). The floating modules use low-voltage high-current silicon transistors (Infineon IPT015N10). These semiconductor devices can for a short time handle more than eight times their continuous data-sheet rating before reaching charge carrier saturation in the channel and terminal damage. Furthermore, these compact transistors allow simple scaling for low loss at rated currents and enough reserve for fault currents. In addition, we increased the thermal capacitance with copper outlays on the circuit board so that the modules can manage typical low-voltage fault currents. We previously tested successful current commutation and operation up to 3,000~A for normal operation of 500~A \cite{GoetzMMC}. Additionally, each transistor costs less than \$4 (further decreasing at high quantities), significantly cheaper than super-junction silicon or even silicon-carbide devices at 1200~V. A three-phase active front end and three single-phase bidirectional dc/dc converters supply the floating modules, and the digital signal processor (Texas Instruments TMS320f28379) controls the setup.
The other parameters are shown in TABLE II.
The results validate the capability of voltage adjustment and phase shift.
\begin{table}
\centering
\caption{Parameters of the experimental setup}
\centering
\begin{tabular}{l c }
\hline
Description & Value\\
\hline
dc link voltage of the floating module & 24~V\\
Switching rate of the floating module & 100~kHz\\
Grid voltage (rms) and frequency & 230~V, 50~Hz\\
Filter inductors & 100~µH~(26 turns) \\
Load & 66$~\Omega$  \\
\hline
\end{tabular}
\label{table:para}
\end{table}

\subsection{Voltage adjustment}

\begin{figure}
    \centering
    \includegraphics[width=0.9\linewidth]{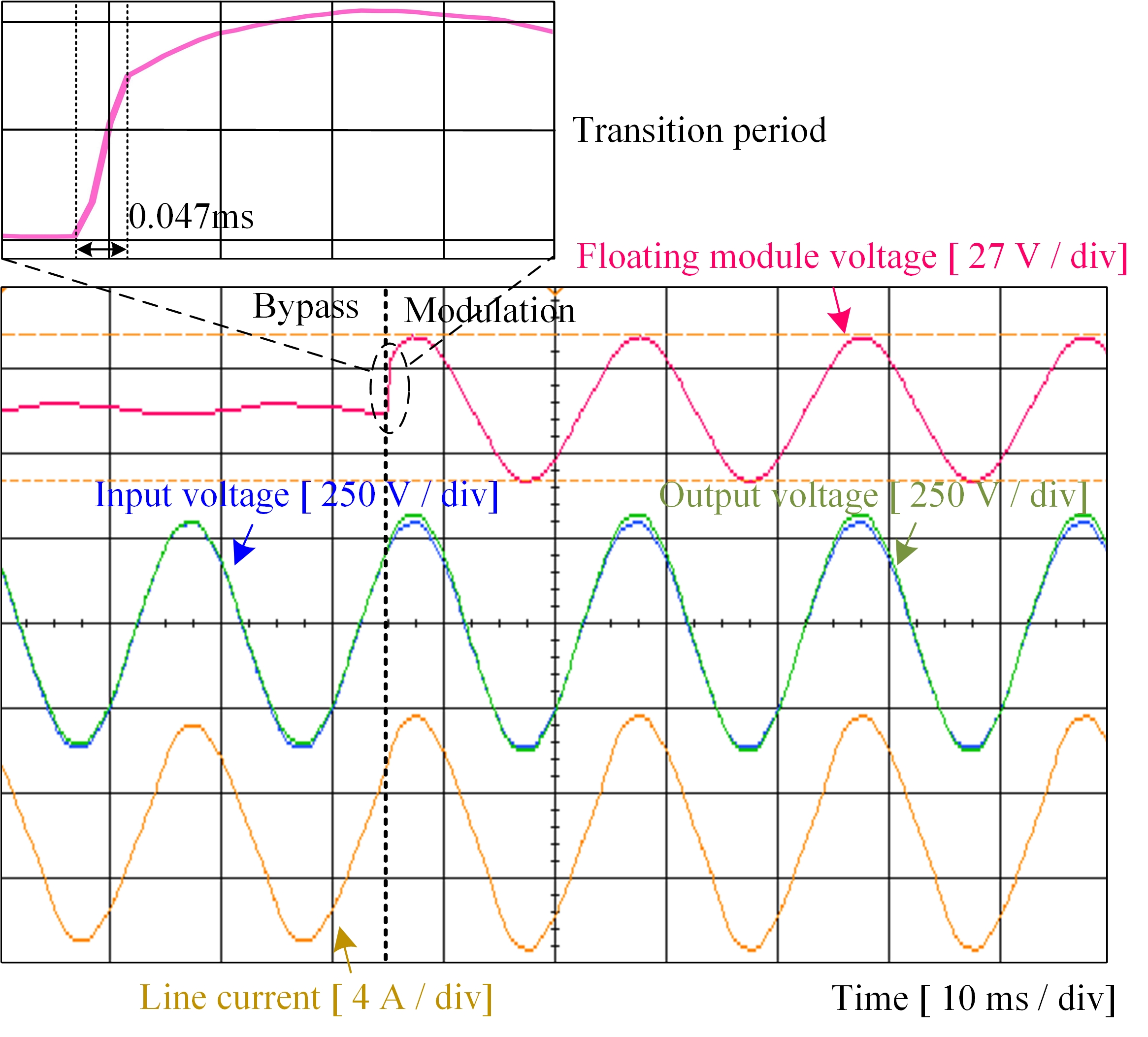}
    \caption{ Experimental results for a voltage increase. }
    \label{fig:increase}
\end{figure}

Fig.~\ref{fig:increase} demonstrates the ability of the floating module to increase the grid voltage. When the circuit changes from the bypass state (the pink curve), which passes the current through without change, to the modulation state, an in-phase voltage is immediately inserted to lift up the voltage (from the blue curve to the green curve), thus increasing the line current (the yellow curve) as well. The inset panel zooms into the transition point where the floating module starts to modulate and demonstrates the fast step response. It takes roughly 47~µs to respond. { When the floating module modulates, the total harmonic distortion (THD) of the grid current is 2.3\%, well below that of the other solutions \cite{Transformer_Less_LED,Hybrid_UPQC}.}

\begin{figure}
    \centering
    \includegraphics[width=0.9\linewidth]{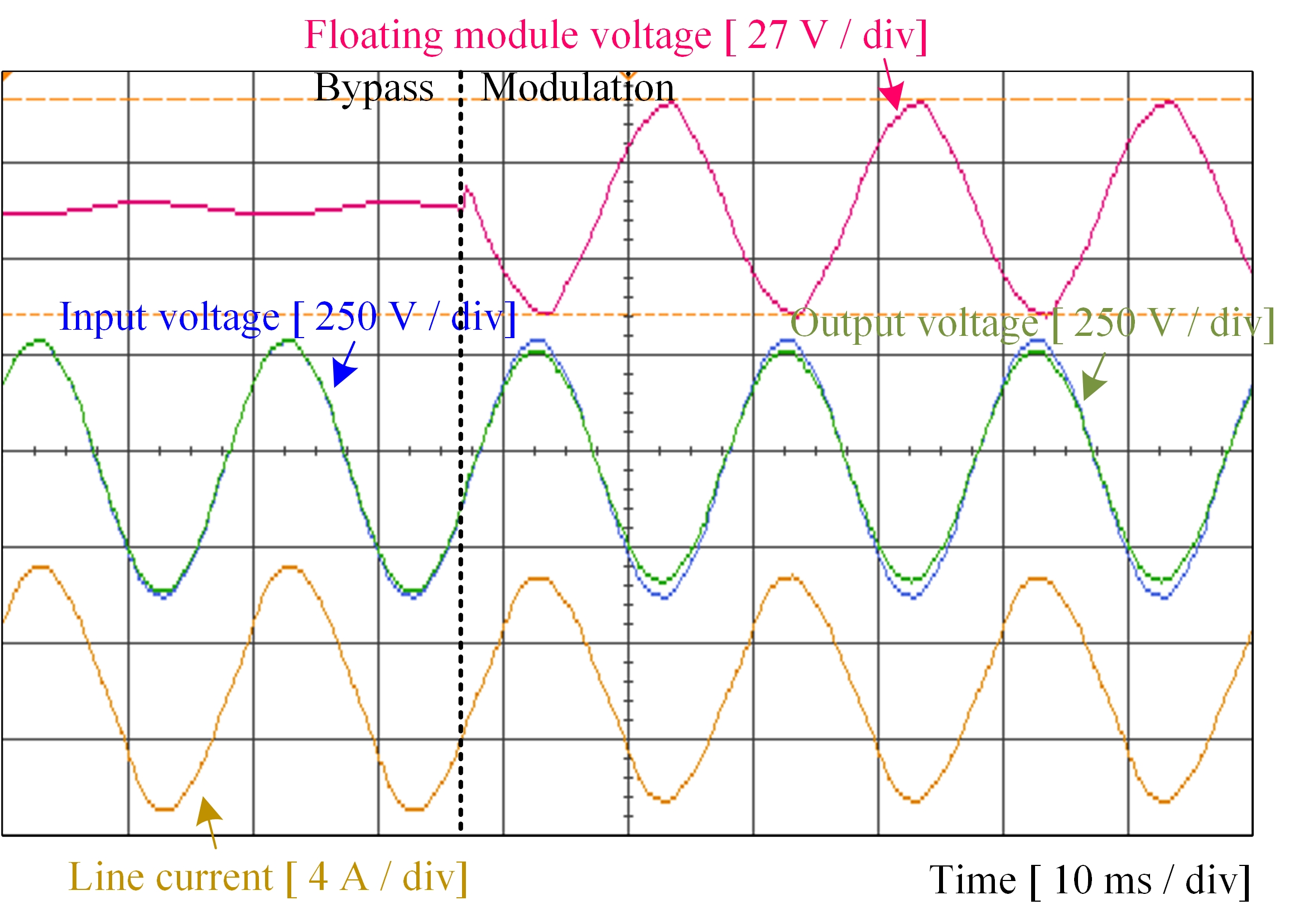}
    \caption{ Experimental results for a voltage decrease. }
    \label{fig:decrease}
\end{figure}

Fig.~\ref{fig:decrease} demonstrates the performance of the floating module in decreasing the grid voltage. When the floating series-injection module starts switching modulation, an inverse voltage is inserted to lower the voltage, thus decreasing the line current as well. 

\subsection{Phase shift}
\begin{figure}
    \centering
    \includegraphics[width=0.9\linewidth]{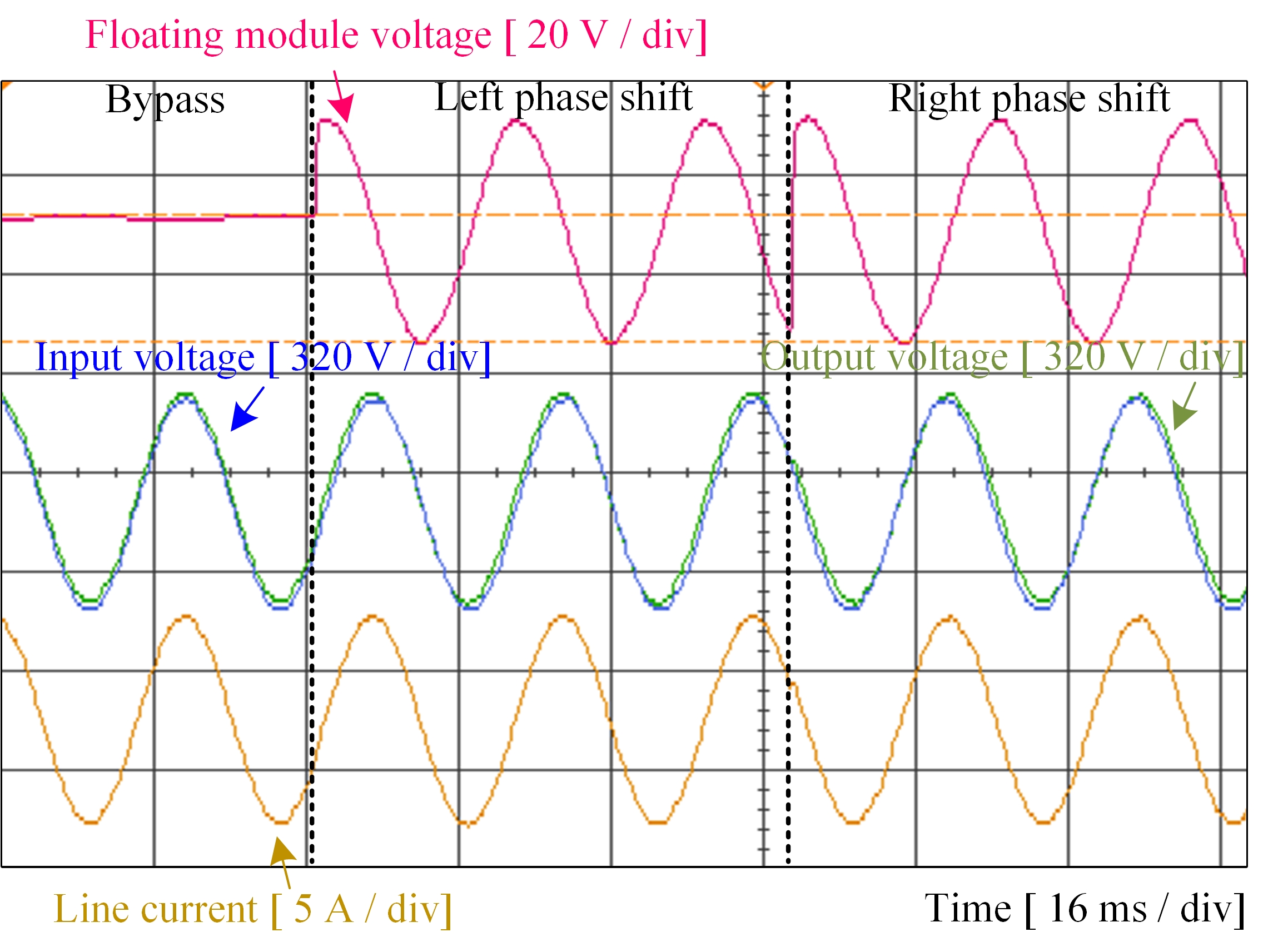}
    \caption{ Experimental results for a phase shift.}
    \label{fig:phase_shift}
\end{figure}

Fig.~\ref{fig:phase_shift} shows the process of shifting the grid voltage ahead or behind (the green curve) of the original one (the blue curve). When a 90° leading voltage (the pink curve) is connected in series with the line, the grid voltage is shifted forward. Subsequently, {the inserted voltage is shifted by 180°}. The now 90° lagging voltage produced by the floating module delays the grid voltage. The direct-injection f/q controller can respond quickly and switch from capacitive to reactive compensation.

\section{Discussion of Practical Aspects}

\subsection{{Size and cost aspects}}
Direct injection with compact high-current lower-voltage semiconductor modules in combination with subjecting all power transfer to high frequencies with a fraction of the power level and smaller magnetics promises size reduction to fit it into typical distribution service cabinets. 

UPFC with injection transformers have the transformer as their weak spot with respect to size as available systems demonstrate. For an example of 500~A rms per phase, the transformer ends up already at a minimum of 80~litres (0.6~kVA/litre for this type of design) \cite{industrialtrans}. The injection transformer is particularly large as it has to deal with the entire line current without saturation, which needs a substantial cross-section of the steel yoke and copper.
Full back-to-back converters, on the other hand, suffer from the size of the converters, which have to deal with the entire line power in both voltage and current. The 500~A example leads to more than 90~litres in total even for optimised commercial air-cooled inverter stacks for the temperature swing in outdoor cabinets with as much as 8~kW/litre \cite{igd4424p1f7blfa,_2023_mtmb6i0300f12inaa}, not yet considering that a back-to-back solution should ideally manage fault currents and needs to be scaled up substantially.

Thus, eliminating line transformers and full-power inverters provides a large volume that can be used for lower-power bidirectional dc/dc converters and the modules. The bidirectional dc/dc converters indeed are rather small. Using the Infineon reference design at 4.2~kW/litre, the dc/dc converters for the example with 500~A line current together need less than twelve litres. The modules are also highly compact, with only three litres. Thus, both the elimination of the injection transformer and the full-power back-to-back converter save volume on such a high level that the system could outcompete them with already a moderately package-optimized implementation.

{ A line-frequency injection transformer at present reaches beyond 250 USD/kW for the relevant power level ($\approx$ 15~kW to 35~kW in this example) due to the tough secondary winding and is strongly driven by rising cost of electrical steel and copper \cite{industrialtrans,4763393,BOULAMANTI2016112}. The previously widely featured SOP solution with a full-power back-to-back converter, on the other hand, generates hardware costs of approximately 2$\times$ 10~USD/kW for industrial  and 2$\times$ 4~USD/kW assuming automotive cost levels \cite{USDRIVERM,semikron}, where the entire line power has to be converted ($\approx$ 150~kW to 350~kW).
At the same time, the specific cost of dc/dc converters reached $<$\,{}50 USD/kW with further decreases expected \cite{USDRIVERM}, and the injection modules at $<$\,{}15~USD/kW, where only a small part of the line power, typically around 10\%, is converted. In consequence, the cost decrease of compact electronics already promises an economic edge of the presented circuit over more conventional solutions, which depend on  the cost of metals as well as resources or are disadvantaged by the high power they need to convert.

}

\subsection{Bandwidth limitation of transformers}
Beyond its size problems, the transformer in conventional UPFC/UPQC also typically restricts the spectral bandwidth and therefore the opportunities for active filtering, which the proposed direct-injection circuit avoids---as do back-to-back converters. Omitting the injection transformer gives access to the full converter bandwidth also for compensating high-order harmonics and transients.
The conventional injection transformer, in contrast, as a side-product of needed high saturation for supporting the full line current and low cost typically demonstrates large eddy current and hysteresis loss already around 1~kHz \cite{The_Influence_of_Complex_Permeability,Comparison_of_lamination_iron}, which further grows as generally known from the reduced Bertotti model per
\begin{equation}
{P_{{\rm{magl}}}} = {P_{{\rm{hys}}}} + {P_{{\rm{edd}}}} = \left( {\eta {B_{\rm{m}}}^2f + \frac{{{\pi ^2}{t^2}{B_{\rm{m}}}^2}}{{6\rho }}{f^2}} \right) V,
\end{equation}
where ${\eta}$ is the Steinmetz constant (roughly 15 here), ${B_{\rm{m}}}$ the flux density, ${f}$ the current frequency, i.e., of a harmonic to be compensated, ${t}$ is the sheet thickness (assuming 0.27~$\upmu$m, which are merely used in power transformers with higher bandwidth already), ${\rho}$ the lamination resistivity (0.48~$\upmu$$\Omega$m), and ${V}$ the volume of the transformer (0.129~m$^{3}$) \cite{_2012_grainoriented}.
Based on the Bertotti model, the transfer function of the transformer-based UPFC/UPQC would follow 
\begin{equation}
H = \left| {\frac{{{P_{\rm{g}}} - {P_{{\rm{magl}}}}}}{{{P_{\rm{g}}}}}} \right|,
\label{equ:gain}
\end{equation}
where ${{P_{\rm{g}}}}$ is the power to be injected to compensate for the harmonic distortion. From Eq.~(\ref{equ:gain}), the 3-dB bandwidth of the conventional transformer-based UPFC is roughly 1~kHz due to damping, whereas direct-injection inverters readily reach ten times that level. 

{ \subsection{Power loss analysis}

Due to more degrees of freedom, channels, and conditions than a normal converter, the loss profile is multi-dimensional. For a case where the phase rms current equals to 100~A, the output voltage of the floating module is 33~V, and the injected power amounts to 3.3~kW, we modeled the proposed circuit and transformer-based UPFC in Plexim PLECS to analyze the power loss. 

For the proposed direct-injection UPFC, we set the parameters based on Infineon IPT015N10 (for the series floating module), Infineon IM828-XCC (for the shunt active front end), and Infineon 3K3W-BIDI-PSFB (for the dc/dc converter). The power loss of the series floating module contributed by switching and conduction amounts to 44.3~W. The power loss of the filtering inductor (5~m$\Omega$) contributes to 50 W. The shunt active front end leads to a power loss of 48.3~W per phase. Meanwhile, each dc/dc converter causes 67~W. 
The total power loss per phase reaches 209.6~W. 

For transformer injection, the shunt and the series part each have power loss comparable to our systems grid interface, i.e., some 48.3~W per phase each. However, the effective injection transformer resistance (e.g., Boardman) is 20 m$\Omega$ \cite{industrialtrans}, leading to 150.3 W per phase for the injection and 246.9 W overall. When injecting a high-order signal through the transformer for harmonic compensation, the loss increases drastically, with the squared frequency as discussed in Section~VI.B.

Overall, the loss of the new circuit is on a similar order of magnitude or better than the prior art with an injection transformer while highly compact and allowing for benefitting from the rapid improvement of power semiconductors in the future (power transistors still follow a Moore-like development), whereas transformers do not develop anymore.

}

\subsection{Comments on Component Selection, Ratings, and Safe State}
The shunt converter, e.g., an active front-end, can widely follow the design rules of other not safety-critical grid-connected inverters, such as power supplies or compensation devices. The safe state if the control detects issues, critical inconsistencies arise, or similar exceptions evolve typically simply turns off the converter by blocking all gate drivers and disconnecting the control signals from the hardware. The availability of the shunt converter is not absolutely necessary.

The modules, however, differ. They are in series with a line and use only limited voltage to maintain differences or control the current flow. In case of an exception, the bypass state of the series modules appears to be the most appropriate response. It avoids disconnecting any loads or grid segments and suppresses the formation of critically high voltage differences. Disconnection of a line, in contrast, rather appears to be the task of a feeder. In case the voltage difference across the system eventually increases, the floating modules may try to compensate the difference through regular modulation. Beyond the standard voltage range, over-modulation allows expanding that range further. If the building-up voltage exceeds even the over-modulation range, the line will have left the acceptable voltage band far behind. Switching the series modules to bypass and let sufficient fault current flow across can reduce the problem and/or trip normal grid-protection measures.

Although this direct-injection power-flow controller could have a mechanical bypass shunt and/or disconnection contactors as conventional solutions use, the high current capability of latest low-voltage transistors allows a compact module design that can encompass grid fault currents to ensure sufficient short-circuit capacitance on both sides of the series modules for voltage stability and reliable triggering of the line protection, typically high-rupturing capacity (HRC) fuses. As however the thermal capacitance of the transistors on their small lead frame is typically notably lower than that of fuses, the modules' $I^2 t$ value with current $I$ and time $t$ would fall short of the fuses' $I^2 t$ if the modules were only designed for the continuous current rating. We for example increased the thermal capacitance in our design with 2.2~mm surface-mount copper outlays (matched to the transistor height) to provide more overload capability. Still, the module rating should ideally reach the line short-circuit current.
A typical 300~kVA transformer with for example 8.5\% short circuit voltage leads to 430~A continuous and 5100~A short-circuit current. A matched HRC fuse would be expected to react within about 400~ms to this current. Latest low-voltage silicon semiconductors from various commercial sources, such as Infineon, ON Semiconductor, or Texas Instruments, allow reaching such short-circuit currents reliably and affordably for the first time.

\section{Conclusions}
{
We presented a fully electronic modular power-flow controller that avoids any bulky grid-frequency transformers and uses semiconductors in the injection modules that may be rated far below the grid voltage. Latest low-voltage semiconductor developments make the topology particularly attractive in low-voltage grids (e.g., 120~V\,{}/\,{}240~V\,{}/\,{}480~V or 230~V\,{}/\,{}400~V), which are currently under enormous pressure due to uncontrolled and often not monitored power flows as well as voltage violations caused by distributed generation and large loads, such as vehicle chargers. 
The principle, however, can be scaled to any voltage level and may provide advantages also in medium and high voltage compared to conventional UPFC designs.

In contrast to other fully electronic power-flow circuit topologies, such as back-to-back converters, it does not need to convert the entire power flowing through a line but only the small difference it injects or extracts.
Despite the limited voltage reserve in the floating modules, even 48~V are sufficient to cover voltage differences of ±15\% in 230~V grids and ±31\% in 110~V grids, while further over-modulation can expand the operating range in case of exceptional events.} The low power requirements are similar to series-transformer injection but without the problems of line-frequency transformers, such as size, cost, loss, and limited frequency bandwidth.

{ The simulation and experimental results validate the key functionalities, e.g., that the compact circuit allows bidirectional active and reactive power-flow control, harmonics injection or blocking, and unbalanced load regulation, with only minor adjustment of control methods known from conventional UPFC and UPQC circuits.}

Flow control has primarily been promoted for high-voltage grids; the massive growth of renewable power generation and large loads in the low-voltage distribution grid, however, lead to uncontrolled power flow in meshed grids and can cause local overload conditions that are not covered by conventional fusing concepts. Whereas tap changers on the level of distribution substations cannot solve such situations, electronic power-flow controllers can. The proposed solution has strong potential to be compact enough to be installed in conventional on-street utility boxes.

 \ifCLASSOPTIONcaptionsoff
  \newpage
\fi

\bibliographystyle{IEEEtran}
\bibliography{reference}

\end{document}